%% file: extension_complexity.tex
\documentclass[11pt]{article}
\usepackage[margin=1in]{geometry}
\usepackage{amsmath,amssymb,amsthm,mathtools}
\usepackage{tikz}
\usepackage[hidelinks]{hyperref}
\hypersetup{pdftitle={Unbounded savings in the extension complexity of Cartesian products},pdfauthor={Minhao Liu}}

\newtheorem{theorem}{Theorem}
\newtheorem{lemma}[theorem]{Lemma}
\newtheorem{proposition}[theorem]{Proposition}
\newtheorem{corollary}[theorem]{Corollary}
\theoremstyle{remark}
\newtheorem{remark}[theorem]{Remark}
\numberwithin{equation}{section}

\newcommand{\xc}{\operatorname{xc}}
\newcommand{\rk}{\operatorname{rank}_+}
\newcommand{\conv}{\operatorname{conv}}
\newcommand{\supp}{\operatorname{supp}}
\newcommand{\st}{\operatorname{st}}
\newcommand{\pyr}{\operatorname{pyr}}
\newcommand{\R}{\mathbb R}
\newcommand{\Z}{\mathbb Z}
\newcommand{\ot}{\mathbin{\otimes_1}}
\newcommand{\repositoryURL}{\href{https://github.com/chineseschrodingercat/product-extension-complexity/tree/99a059b98d06e1b912bb033b56314121024c9040}{\texttt{github.com/}\allowbreak\texttt{chineseschrodingercat/}\allowbreak\texttt{product-}\allowbreak\texttt{extension-}\allowbreak\texttt{complexity}} (commit \texttt{99a059b98d06})}
\title{Unbounded savings in the extension complexity\\of Cartesian products}
\author{Minhao Liu\\[3pt]\small Texas Tech University, Lubbock, TX 79409, USA\\\small\href{mailto:minhliu@ttu.edu}{minhliu@ttu.edu}}
\date{}

\begin{document}
\maketitle

\begin{abstract}
The extension complexity of a Cartesian product of polytopes is at most the sum of the extension complexities of its factors, and Grande, Padrol and Sanyal conjectured that equality always holds. A counterexample, a polytope whose Cartesian square saves at least one inequality, was given earlier in a manuscript released by Shitov. The results of this paper were obtained independently, and they show that the saving is unbounded. For every positive integer $t$, we construct a rational polytope $R_t$ whose Cartesian square needs at least $t$ fewer inequalities in a linear extended formulation than two optimal formulations of $R_t$ taken separately. For a family of pairs $(U_n,V_n)$ of different rational polytopes, the separate extension complexities sum to $O(n)$ and the product saves at least $\lceil5n/27\rceil$ inequalities. The saving is therefore at least a fixed positive fraction of the total. We also give a $24$-dimensional rational polytope of extension complexity $40$ whose square has extension complexity exactly $79$. A transfer formula for facet wedges expresses the saving in terms of nonnegative ranks of partial slack matrices, and it shows that additivity for all polytopes is equivalent to additivity for these matrices. For squares, intersecting families of sets on successively larger alphabets let the two copies share generators while forcing each coordinate to contribute its full cost within one factor.
\end{abstract}


\section{Introduction}

The \emph{linear extension complexity} $\xc(P)$ of a polytope $P$ is the minimum number of inequalities in a linear description of a polyhedron that projects onto $P$, with equations counted for free. Extended formulations are used throughout combinatorial optimization \cite{ccz,kaibel}, and their sizes have also been studied for general polytopes \cite{ksz}. Placing optimal formulations of $U$ and $V$ side by side gives a formulation of $U\times V$ with $\xc(U)+\xc(V)$ inequalities. The number saved by an optimal joint formulation is the \emph{deficit}
\[
  \delta(U,V)=\xc(U)+\xc(V)-\xc(U\times V).
\]

The question whether this deficit always vanishes arose at the 2013 Dagstuhl seminar on communication complexity and linear optimization \cite{dagstuhl,gps}. It appears as Problem~3 in Weltge's dissertation \cite{weltge} and as Conjecture~1.2 of Grande, Padrol and Sanyal \cite{gps}, motivated by the fact that a direct sum of matroids gives a Cartesian product of their base polytopes. Additivity holds when one factor is a simplex \cite[Corollary~2.4]{gps} and, more generally, when one factor is a pyramid \cite{twz}. Weltge posed the general question again at the 2022 Carg\`ese workshop \cite{cargese}.

The conjecture is false. A counterexample was made public by Shitov in early November 2026 \cite[Section~6.1]{shitovoverview}. The proof is in manuscript~6.1, dated 2~November 2026, of the archive distributed with \cite{shitovoverview}; we refer to it as \cite[manuscript~6.1]{shitovoverview}. Its Theorem~1 constructs a $59$-dimensional polytope $H$ with $\xc(H)=92$ and $\xc(H\times H)\le183$, so that $\delta(H,H)\ge1$. The first version of the present paper (\href{https://arxiv.org/abs/2611.33631v1}{arXiv:2611.33631v1}) was posted about three weeks later and was written without knowledge of \cite{shitovoverview}, which therefore has priority for the disproof of the conjecture; I learned of it from Y.~Shitov after that version had appeared. All results of this paper, including a second counterexample (Theorem~\ref{thm:main}), were obtained independently, and their statements and proofs are unchanged from the first version. The two approaches share two ingredients, which we point out where they occur: a set of three labels of which every input admits at least two, and a bound for splitting a nonnegative quantity into two parts.

How large can the saving be? Pairing $2m$ copies of $H$ gives $\xc(H^{2m})\le183m$, which saves $m$ inequalities relative to $2m$ separately optimized copies. This does not show that the deficit can be large. To estimate the deficit of the square of $H^m$, one must compare this cost with $2\xc(H^m)$, and the $m$ copies within each factor may already share inequalities. A large deficit therefore requires lower bounds on the individual optima. Our main theorem bounds the joint cost from above and the two individual optima from below.

\begin{theorem}\label{thm:binary}
For every integer $n\ge1$ there are positive-dimensional rational polytopes $U_n,V_n$ with
\[
                    \delta(U_n,V_n)\ge\left\lceil 5n/27\right\rceil .
\]
Their separate complexities sum to at most $Kn$ for an absolute constant $K$. In particular, there is an $\eta>0$, independent of $n$, such that $\xc(U_n\times V_n)\le(1-\eta)(\xc(U_n)+\xc(V_n))$.

For every integer $t\ge1$ there is also a positive-dimensional rational polytope $R_t$ such that
\[
                      \delta(R_t,R_t)\ge t .
\]
\end{theorem}

We first construct nonnegative matrices. The nonnegative rank of such a matrix is the least number of nonnegative vectors, called generators, whose cone contains its columns. The slack matrix of a Cartesian product stacks one column from each factor for every pair of columns. For the factors $U_n$ and $V_n$ we use four labels in each of $n$ coordinates. A column in the first matrix admits two labels in a coordinate, and a column in the second admits three. The two sets always meet, so four shared generators suffice per coordinate in the product, in addition to the generators used on one side only. Within a single factor, one generator may serve several coordinates. We fix the choices in all but one coordinate and average the resulting lower bounds over those choices. This gives the proportional saving in Theorem~\ref{thm:binary}.

For the squares, coordinate $i$ uses all $j_i$-element subsets of a $(2j_i-1)$-element set. Any two such subsets meet, whereas a set meeting every one of them has at least $j_i$ elements. If the alphabets grow sufficiently fast, each coordinate incurs its full cost even after the previous coordinates have been optimized together (Lemma~\ref{lem:largealphabet}). The joint formulation then saves at least one generator per coordinate.

We also give an example in which both the individual and the joint optimum are determined exactly; for the polytope $H$ of \cite[manuscript~6.1]{shitovoverview}, the complexity of the square is bounded from above.

\begin{theorem}\label{thm:main}
There is a rational polytope $P$ of dimension $24$ with $50$ facets and $150$ vertices such that
\[
  \xc(P)=40\qquad\text{and}\qquad\xc(P\times P)=79 .
\]
\end{theorem}

The $50$ integer inequalities in~\eqref{eq:P} describe $P$ explicitly. The saving has the same source as in \cite[manuscript~6.1, Lemma~2 and Proposition~3]{shitovoverview}, where every point can be lifted using at least two of the three vertices of an auxiliary triangle, so that any two points can use a common vertex. Here a matrix $S$ formed by $24$ facet rows has twelve fixed star vectors and three further vectors $A_0,A_1,A_2$, and each column can use either of two of the latter. Two of the three vectors suffice for one copy of $S$, and the three paired vectors $(A_s,A_s)$ suffice for two copies, because any two $2$-element subsets of a $3$-element set intersect. This gives the upper bounds $14$ and $27$ for the nonnegative ranks of $S$ and of the corresponding matrix for two copies. The main difficulty is to prove that these bounds are optimal. The ordinary rank and the rectangle covering bound are at most $12$ for one copy and at most $24$ for two (Remark~\ref{rem:standard}), so the proof uses the magnitudes of the entries. A large parameter forces generators close to the stars, and an alternating sum around a six-cycle detects the additional generators that are needed (Proposition~\ref{prop:ranks}).

These matrix constructions give polytopes by a common transfer formula (Theorem~\ref{thm:transfer}). Starting with selected rows of a slack matrix, we wedge over the omitted facets. Each wedge contributes exactly two to the extension complexity in addition to the nonnegative rank of the retained rows. The same formula holds on the product, so the wedge contributions cancel in the deficit. Consequently, additivity for all polytopes is equivalent to additivity for the matrices formed by selected facet rows of slack matrices (Corollary~\ref{cor:equivalence}). The proof rests on a splitting bound (Lemma~\ref{lem:split}). The same bound appears, for families of nonnegative functions on an arbitrary domain, as \cite[manuscript~6.1, Lemma~6]{shitovoverview}.

Strict savings also occur for tensor rank under direct sums \cite{shitov} and for positive semidefinite rank under Cartesian products \cite{grt}. For nonnegative rank, every nonzero rank-one summand of a block-diagonal matrix is supported on a single diagonal block, which gives additivity on block-diagonal sums. The slack matrix of a Cartesian product pairs a column from each factor, allowing a single generator to serve both.

Section~\ref{sec:transfer} proves the transfer formula. Sections~\ref{sec:choices}--\ref{sec:geometry} give the exact example of Theorem~\ref{thm:main}. Section~\ref{sec:binary} proves Theorem~\ref{thm:binary}, first for two different factors and then for Cartesian squares. Section~\ref{sec:discussion} gives consequences for polytope operations and zero-error channel simulation, and states the open questions.

\section{A transfer theorem for partial slack matrices}\label{sec:transfer}

We begin with notation. For a nonnegative matrix $S$, the nonnegative rank $\rk S$ is the smallest number of nonnegative vectors whose conic hull contains all columns of $S$; all factorizations are over the reals. For nonnegative matrices $S\in\R^{m\times n}$ and $T\in\R^{m'\times n'}$, the \emph{$1$-product} $S\ot T\in\R^{(m+m')\times nn'}$ is the matrix whose columns are the vectors $(S_i,T_j)$, where $S_i$ and $T_j$ run over the columns of $S$ and $T$ \cite{aprile}. Here and below, $(x,x')$ denotes the column vector obtained by stacking $x$ on top of $x'$; thus the rows of $S\ot T$ are the rows of $S$ followed by the rows of $T$.

Let $Q$ be a polytope of positive dimension. We write its facet inequalities as $\sigma_f(x)\ge0$ with affine functions $\sigma_f$, one for each facet $f$. The slack matrix $S_Q$ has rows indexed by the facets and columns indexed by the vertices of $Q$, with entries $\sigma_f(v)$; rescaling rows by positive factors does not change $\rk S_Q$. Yannakakis' factorization theorem \cite{yannakakis} states that $\xc(Q)=\rk S_Q$. The facets of $Q\times Q'$ are the products of a facet of one factor with the other factor, and its vertices are the pairs of vertices, so the slack matrix of $Q\times Q'$ is $S_Q\ot S_{Q'}$.

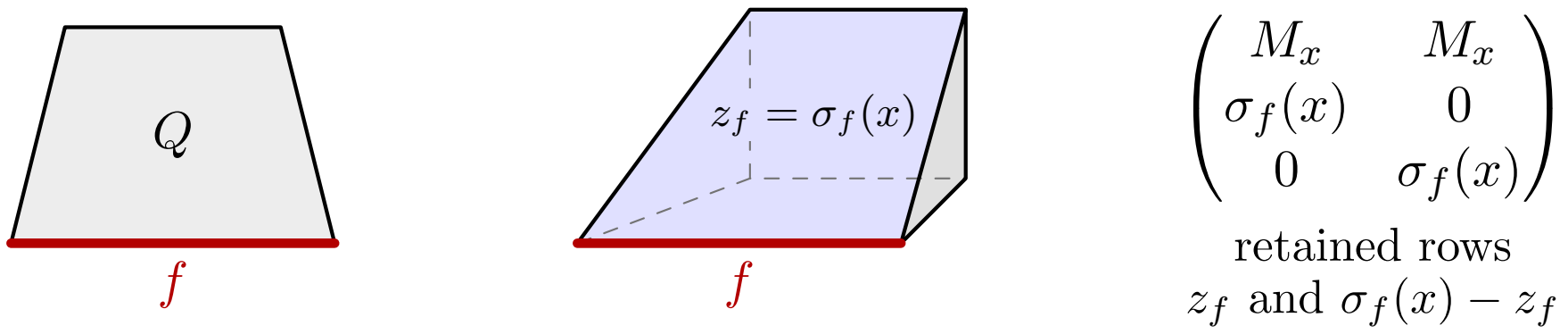
\begin{figure}[htb]
\centering
\begin{tikzpicture}[line join=round,line cap=round]
  \begin{scope}
    \coordinate (a1) at (0,0); \coordinate (a2) at (2.4,0); \coordinate (a3) at (2.0,1.6); \coordinate (a4) at (0.4,1.6);
    \fill[black!7] (a1)--(a2)--(a3)--(a4)--cycle;
    \draw[thick] (a2)--(a3)--(a4)--(a1);
    \draw[line width=2pt,red!70!black] (a1)--(a2);
    \node[below,red!70!black] at (1.2,0) {$f$};
    \node at (1.2,0.8) {$Q$};
    \node[font=\small] at (1.2,-1.05) {(a)};
  \end{scope}
  \begin{scope}[xshift=4.2cm,x={(1cm,0cm)},y={(0.55cm,0.3cm)},z={(0cm,1cm)}]
    \coordinate (b1) at (0,0,0);   \coordinate (b2) at (2.4,0,0);
    \coordinate (b3) at (2.0,1.6,0); \coordinate (b4) at (0.4,1.6,0);
    \coordinate (t3) at (2.0,1.6,1.25); \coordinate (t4) at (0.4,1.6,1.25);
    \fill[black!12] (b2)--(b3)--(t3)--cycle;
    \fill[blue!12] (b1)--(b2)--(t3)--(t4)--cycle;
    \draw[dashed,black!55] (b1)--(b4)--(b3); \draw[dashed,black!55] (b4)--(t4);
    \draw[thick] (b2)--(b3)--(t3)--(b2);
    \draw[thick] (t3)--(t4)--(b1);
    \draw[line width=2pt,red!70!black] (b1)--(b2);
    \node[below,red!70!black] at (1.2,0,0) {$f$};
    \node[font=\footnotesize,fill=blue!12,inner sep=1pt] at (1.27,0.88,0.686) {$z_f=\sigma_f(x)$};
    \node[font=\small] at (1.6,0,-1.05) {(b)};
  \end{scope}
  \begin{scope}[xshift=10.1cm]
    \node at (0,1.0) {$\begin{pmatrix} M_x & M_x\\ \sigma_f(x) & 0\\ 0 & \sigma_f(x)\end{pmatrix}$};
    \node[font=\small,align=center] at (0,-0.25) {retained rows\\[-1pt] $z_f$ and $\sigma_f(x)-z_f$};
    \node[font=\small] at (0,-1.05) {(c)};
  \end{scope}
\end{tikzpicture}
\caption{A wedge over one facet. (a) A polygon $Q$ and a facet $f$ that is not retained. (b) The wedge $\widehat Q=\{(x,z_f):x\in Q,\ 0\le z_f\le\sigma_f(x)\}$. The facet $f$ becomes the edge where the hidden bottom $z_f=0$ meets the top $z_f=\sigma_f(x)$, while the other facets of $Q$ lift to facets of $\widehat Q$. (c) A vertex $x$ of $Q$ with $\sigma_f(x)>0$ has two lifts. On them the retained rows repeat the column $M_x$, and the two new slack rows take the values $\sigma_f(x)e_1$ and $\sigma_f(x)e_2$. This is the matrix $M^\sigma$ of Lemma~\ref{lem:split}, which raises the nonnegative rank by exactly two.}
\label{fig:wedge}
\end{figure}

Now we select some facets of $Q$ to retain and let $M$ be the submatrix of $S_Q$ formed by their rows, with all vertices of $Q$ as columns. For each of the remaining $k$ facets $f$ we introduce a new variable $z_f$ and define
\begin{equation}\label{eq:wedge}
  \widehat Q=\bigl\{(x,z):\ \sigma_f(x)\ge0\ \text{for retained } f,\ \ 0\le z_f\le\sigma_f(x)\ \text{for omitted } f\bigr\}.
\end{equation}
The omitted inequalities $\sigma_f(x)\ge0$ are implied, so $\widehat Q$ projects onto $Q$. Geometrically, $\widehat Q$ arises from $Q$ by wedging successively over the $k$ omitted facets; Figure~\ref{fig:wedge} shows a single wedge.

\begin{theorem}[Transfer by facet wedges]\label{thm:transfer}
Let $Q$ and $Q'$ be polytopes of positive dimension, and let $M$ and $M'$ be the submatrices of $S_Q$ and $S_{Q'}$ formed by selected facet rows, with $k$ and $k'$ facets omitted, respectively. Then
\[
  \xc(\widehat Q)=\rk M+2k
  \qquad\text{and}\qquad
  \xc(\widehat Q\times\widehat Q')=\rk(M\ot M')+2k+2k' .
\]
In particular,
\[
  \xc(\widehat Q)+\xc(\widehat Q')-\xc(\widehat Q\times\widehat Q')=\rk M+\rk M'-\rk(M\ot M').
\]
\end{theorem}

Here a matrix without rows has nonnegative rank zero. Adding columns that are convex combinations of other columns changes neither side of these formulas, because the corresponding columns of $M\ot M'$ are again convex combinations. The theorem thus turns the question about polytopes into a question about matrices.

\begin{corollary}\label{cor:equivalence}
The equality $\xc(Q\times Q')=\xc(Q)+\xc(Q')$ holds for all polytopes $Q$ and $Q'$ if and only if $\rk(M\ot M')=\rk M+\rk M'$ for all submatrices $M$ and $M'$ of slack matrices of polytopes of positive dimension formed by selected facet rows.
\end{corollary}

\begin{proof}
If the matrix identity holds, we apply it with $M=S_Q$ and $M'=S_{Q'}$ and use Yannakakis' theorem; polytopes of dimension zero are trivial. Conversely, if additivity holds for all polytopes, it holds for $\widehat Q$ and $\widehat Q'$, and Theorem~\ref{thm:transfer} yields the matrix identity.
\end{proof}

Some restriction on the matrices is necessary, as the example $[1]\ot[1]$ shows. The restriction comes from the zero pattern of slack matrices: every facet row vanishes at the vertices of its facet. The following lemma isolates this property and describes the effect of a single wedge at the level of matrices. Its lower bound is the case of finitely many columns of \cite[manuscript~6.1, Lemma~6]{shitovoverview}. That result allows all nonnegative splits of a prescribed value into two parts. For each column, those splits form the segment between the two endpoints used below, and a convex cone contains the segment exactly when it contains both endpoints. The two proofs were found independently and use the same idea. We include ours to keep the paper self-contained.

\begin{lemma}\label{lem:split}
Let $M$ be a nonnegative matrix with columns $M_1,\dots,M_n$, and let $\sigma\in\R^n_{\ge0}$ be a nonzero row vector. Suppose that for every nonzero row $g$ of $M$ there is an index $i$ with $M_{gi}=0<\sigma_i$. Let $M^\sigma$ be the matrix with the $2n$ columns $(M_i,\sigma_ie_t)$, where $1\le i\le n$, $t\in\{1,2\}$, and $e_1,e_2$ are the unit vectors of $\R^2$. Then $\rk M^\sigma=\rk M+2$.
\end{lemma}

\begin{proof}
Adding the vectors $(0,e_1)$ and $(0,e_2)$ to a generating set of $M$ of minimum size shows that $\rk M^\sigma\le\rk M+2$. Conversely, suppose that the columns of $M^\sigma$ lie in the cone of $r\le\rk M+1$ nonzero nonnegative vectors. We fix a representation of every column and discard the vectors that are used in none of them. For $t=1,2$, let $G_t$ be the set of remaining vectors whose coordinate in the bottom row $t$ is positive. Since $\sigma\neq0$, both sets are nonempty. They are disjoint, because a vector in $G_1\cap G_2$ is used by some column, which would then have two positive bottom coordinates.

Consider the columns with $t=1$. Their bottom row $2$ vanishes, so the vectors in $G_2$ have coefficient zero in them, and the upper parts of the vectors outside $G_2$ generate $M$. Hence $r-|G_2|\ge\rk M$, that is, $|G_2|\le1$; by symmetry also $|G_1|\le1$. We write $G_t=\{h_t\}$. If $\sigma_i>0$, the column $(M_i,\sigma_ie_t)$ uses $h_t$ with a positive coefficient, because $h_t$ is the only vector that is positive in the bottom row $t$. If $g$ is a nonzero row of $M$ and $M_{gi}=0<\sigma_i$, this column vanishes in row $g$, and therefore so does $h_t$. Zero rows of $M$ vanish in every vector that is used. Thus $h_1$ and $h_2$ vanish on the whole upper block, and the remaining $r-2\le\rk M-1$ vectors generate $M$, a contradiction.
\end{proof}

\begin{proof}[Proof of Theorem~\ref{thm:transfer}]
We proceed in three steps.

\emph{Step 1: vertices and facets of $\widehat Q$.} For a vertex $x$ of $Q$ and a choice of an endpoint $z_f\in\{0,\sigma_f(x)\}$ for every omitted facet $f$, we call $(x,z)$ a \emph{lift} of $x$. We claim that the vertices of $\widehat Q$ are exactly the lifts of the vertices of $Q$. If a point $(x,z)$ of $\widehat Q$ satisfies $0<z_f<\sigma_f(x)$ for some $f$, we can move $z_f$ in both directions, so $(x,z)$ is not a vertex. Every vertex of $\widehat Q$ therefore chooses either the lower or the upper endpoint at every omitted facet. Fixing one endpoint choice for each omitted facet cuts out a face of $\widehat Q$ that is the image of $Q$ under an affine isomorphism $x\mapsto(x,z(x))$. Its vertices are the corresponding lifts of the vertices of $Q$: as vertices of a face they are vertices of $\widehat Q$, and conversely every vertex of $\widehat Q$ lies in one of these faces and is therefore a vertex of it, hence a lift. If $\sigma_f(x)=0$, the two choices at $f$ give the same lift, which merely duplicates a column of the slack matrix. Over an interior point of $Q$ all $k$ intervals are nondegenerate, so $\dim\widehat Q=\dim Q+k$. Each retained inequality defines a facet: we take $x$ in the relative interior of the facet, where every omitted slack is positive, and choose every $z_f$ strictly inside its interval. Each new inequality defines a facet as well: we take an interior point $x$ of $Q$, put $z_f$ at the chosen endpoint and all other new coordinates strictly inside their intervals. In both cases only the inequality under consideration is tight.

\emph{Step 2: removing the wedges.} The slack matrix of $\widehat Q$ consists of the rows of $M$, evaluated at the lifts, and the two rows $z_f$ and $\sigma_f(x)-z_f$ for every omitted facet $f$. At a lift, this pair of rows equals $\sigma_f(x)e_1$ or $\sigma_f(x)e_2$, depending on the chosen endpoint. We order the omitted facets as $f_1,\dots,f_k$. Deleting the two rows of $f_k$, and identifying lifts that agree at $f_1,\dots,f_{k-1}$, leaves the matrix formed by the rows of $M$ and the rows of the first $k-1$ wedges, evaluated at the lifts for these wedges. The full matrix is recovered from it by the construction of Lemma~\ref{lem:split} with $\sigma=\sigma_{f_k}$. We verify the hypothesis of the lemma. A row of a retained facet $F$ vanishes at every vertex of $F$; since distinct facets are incomparable, $F$ has a vertex $x$ with $\sigma_{f_k}(x)>0$, and every lift of $x$ will do. The row $z_{f_j}$ (respectively $\sigma_{f_j}-z_{f_j}$) of an earlier wedge vanishes at every lift whose choice at $f_j$ is the lower (respectively upper) endpoint, and we take such a lift of any vertex $x$ with $\sigma_{f_k}(x)>0$. Hence each removal lowers the nonnegative rank by exactly two. After $k$ removals we arrive at $M$, and Yannakakis' theorem gives $\xc(\widehat Q)=\rk M+2k$.

\emph{Step 3: products.} The slack matrix of $\widehat Q\times\widehat Q'$ is the $1$-product of the slack matrices of $\widehat Q$ and $\widehat Q'$. We remove the $k+k'$ pairs of wedge rows one at a time, as in Step~2, and verify the hypothesis of Lemma~\ref{lem:split} at each removal. Suppose that we remove a wedge of $\widehat Q$ with slack $\sigma$, which depends only on the first component of a column. For a row belonging to $\widehat Q$ we choose the first component as in Step~2 and the second component arbitrarily. For a row belonging to $\widehat Q'$ we choose a first component with $\sigma>0$ and a second component at which the row vanishes, namely a vertex of the corresponding facet of $Q'$ or a lift with a suitable endpoint. The wedges of $\widehat Q'$ are removed symmetrically. After all removals we arrive at $M\ot M'$, and Yannakakis' theorem gives the second formula.
\end{proof}

\begin{remark}\label{rem:wedge}
For a single omitted facet $F$, Theorem~\ref{thm:transfer} describes the wedge of $Q$ over $F$: its extension complexity is $\rk S_Q^{-F}+2$, where $S_Q^{-F}$ denotes $S_Q$ with the row of $F$ deleted. Deleting a row lowers the nonnegative rank by at most one, so a wedge raises the extension complexity by one or two. By comparison, a pyramid always raises it by exactly one \cite{padrol}. Since the wedge is polar to the one-point suspension, the same formula holds for one-point suspensions, with a column of the slack matrix deleted instead of a row. Padrol uses one-point suspensions in the classification of polytopes with few vertices \cite{padrol}.
\end{remark}

\section{Three choices on a six-cycle}\label{sec:choices}

We now construct the matrix $S$. Let $G$ be the graph obtained from a hexagon by replacing each vertex $j\in\Z/6$ with a pair $C_j$ of twin nodes and joining every node of $C_j$ to both nodes of $C_{j+1}$. We call $C_j$ a \emph{cluster}, and we call the four edges between $C_j$ and $C_{j+1}$ the edges of \emph{type}~$j$. Thus $G$ has $12$ nodes and $24$ edges, and it is bipartite, with the nodes of the even clusters $C_0,C_2,C_4$ on one side. Let $R\in\{0,1\}^{24\times12}$ be the unsigned edge--node incidence matrix of $G$, and let $R_v$ be its column at the node $v$. The support of $R_v$ is the \emph{star} $\st(v)$, which consists of the four edges at $v$. Two distinct stars share at most one edge.

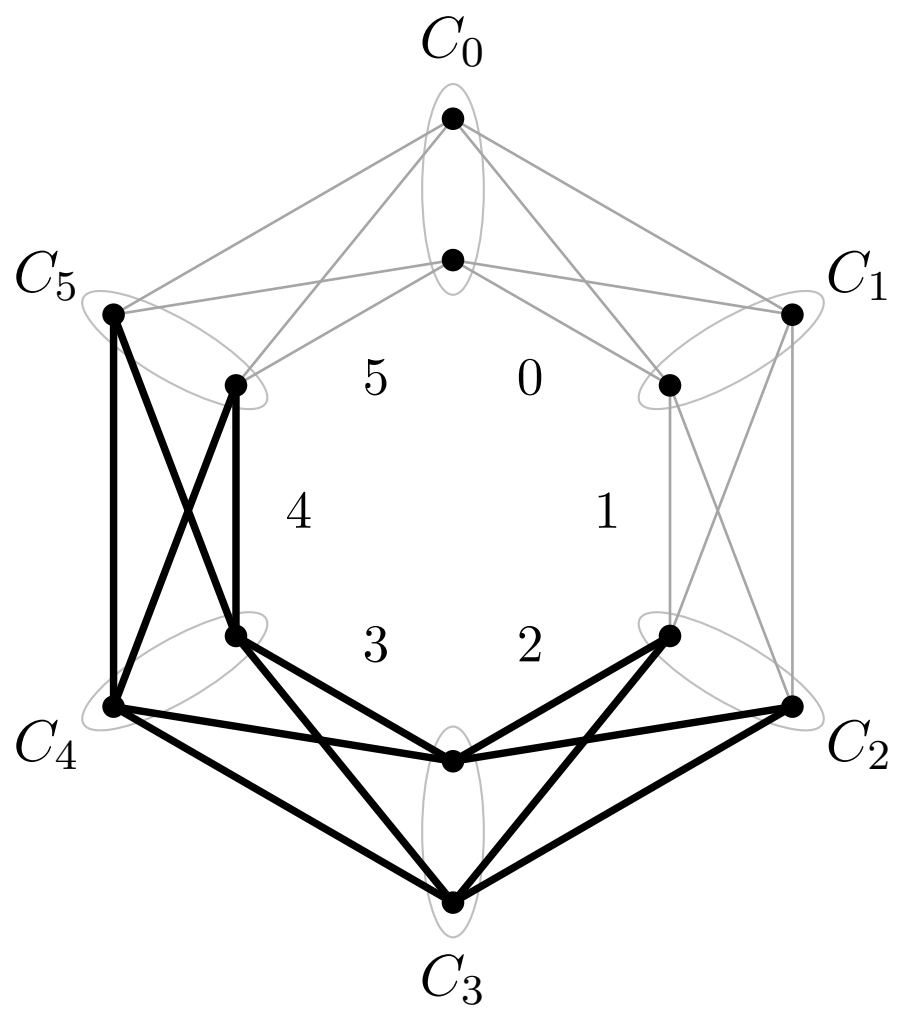
\begin{figure}[htb]
\centering
\begin{tikzpicture}[scale=0.85,
    vtx/.style={circle,fill=black,inner sep=1.5pt},
    light/.style={black!35,line width=0.5pt},
    heavy/.style={black,line width=1.4pt}]
  \foreach \j in {0,...,5}{
    \pgfmathsetmacro{\aj}{90-60*\j}
    \pgfmathsetmacro{\mj}{60-60*\j}
    \coordinate (a\j) at (\aj:1.95);
    \coordinate (b\j) at (\aj:3.05);
    \coordinate (c\j) at (\aj:2.5);
    \draw[black!25,rotate around={\aj:(c\j)}] (c\j) ellipse (0.82 and 0.24);
    \node at (\aj:3.65) {$C_{\j}$};
    \node[font=\small] at (\mj:1.2) {$\j$};
  }
  \foreach \j/\k in {5/0,0/1,1/2}{
    \foreach \p in {a,b}{\foreach \q in {a,b}{\draw[light] (\p\j)--(\q\k);}}}
  \foreach \j/\k in {2/3,3/4,4/5}{
    \foreach \p in {a,b}{\foreach \q in {a,b}{\draw[heavy] (\p\j)--(\q\k);}}}
  \foreach \j in {0,...,5}{\node[vtx] at (a\j){}; \node[vtx] at (b\j){};}
\end{tikzpicture}
\caption{The graph $G$. Each cluster $C_j$ consists of two twin nodes, and the four edges between $C_j$ and $C_{j+1}$ have type $j$ (numbers inside the hexagon). The heavy edges form the support of $X_0$, which is the union of the stars of the four nodes in $C_3\cup C_4$. The vectors $A_0,A_1,A_2$ are the indicator vectors of the edge types $0,2,4$. The two representations $X_0=A_1+\sum_{v\in C_4}R_v=A_2+\sum_{v\in C_3}R_v$ in~\eqref{eq:two} can be read off from the picture.}
\label{fig:graph}
\end{figure}

For $s\in\Z/3$ let $A_s\in\{0,1\}^{24}$ be the indicator vector of the edges of type $2s$, and for $t\in\Z/3$ let $X_t$ be the indicator vector of the edges of the three consecutive types $2t+2$, $2t+3$ and $2t+4$, with indices modulo~$6$; see Figure~\ref{fig:graph}. We fix $L=10^6$ and let $S$ be the $24\times39$ matrix with the columns
\begin{equation}\label{eq:columns}
  X_t\qquad\text{and}\qquad X_t+LR_v\qquad(t\in\Z/3,\ v\in V(G)).
\end{equation}

The next lemma records the two ways of generating each column and the resulting upper bounds. Any two columns have a common admissible vector $A_s$; the auxiliary triangle of \cite[manuscript~6.1, Lemma~2]{shitovoverview} has the same property.

\begin{lemma}\label{lem:upper}
For every $t\in\Z/3$,
\begin{equation}\label{eq:two}
  X_t=A_{t+1}+\sum_{v\in C_{2t+4}}R_v=A_{t+2}+\sum_{v\in C_{2t+3}}R_v .
\end{equation}
Consequently $\rk S\le14$ and $\rk(S\ot S)\le27$.
\end{lemma}

\begin{proof}
The stars of the two nodes of $C_{2t+4}$ cover every edge of the types $2t+3$ and $2t+4$ exactly once, and $A_{t+1}$ adds the edges of type $2t+2$; this proves the first identity. Similarly, the stars of the two nodes of $C_{2t+3}$ cover the types $2t+2$ and $2t+3$, and $A_{t+2}$ adds type $2t+4$. Adding $LR_v$ to both sides gives the same two representations of $X_t+LR_v$. Hence every column of $S$ with index $t$ is a nonnegative combination of the twelve stars and of $A_s$, with coefficient one on $A_s$, for each $s$ in the two-element set $\Z/3\setminus\{t\}$.

Since $\{0,1\}$ meets each of the three sets $\Z/3\setminus\{t\}$, the twelve stars together with $A_0$ and $A_1$ generate $S$, and $\rk S\le14$. For the $1$-product, we consider two columns of $S$ with indices $t$ and $t'$. The sets $\Z/3\setminus\{t\}$ and $\Z/3\setminus\{t'\}$ intersect, and for $s$ in the intersection we write the pair of columns as a nonnegative combination of the $27$ vectors
\begin{equation}\label{eq:productdict}
  (R_v,0),\qquad(0,R_v),\qquad(A_s,A_s)\qquad(v\in V(G),\ s\in\Z/3),
\end{equation}
with coefficient one on $(A_s,A_s)$. Hence $\rk(S\ot S)\le27$.
\end{proof}

The saving in Lemma~\ref{lem:upper} comes entirely from the shared vectors $(A_s,A_s)$. We now show that it is genuine: neither upper bound can be improved.

\section{The matching lower bounds}\label{sec:lower}

The proof rests on two properties of the construction. First, we choose one node in each cluster $C_j$. These six nodes span a six-cycle in $G$ that contains exactly one edge $e_j$ of each type $j$, and we consider the alternating sum
\[
  D(x)=\sum_{j=0}^{5}(-1)^j x_{e_j}\qquad(x\in\R^{24}).
\]
It satisfies
\begin{equation}\label{eq:cycle}
  D(R_v)=0\quad(v\in V(G)),\qquad D(A_s)=D(X_t)=1\quad(s,t\in\Z/3).
\end{equation}
Indeed, if $v$ lies on the cycle, its star contains exactly the two cycle edges at $v$, and these carry opposite signs; otherwise its star contains no cycle edge at all. The support of $A_s$ is a single even type, and the support of $X_t$ consists of three consecutive types beginning with the even type $2t+2$, so the signs are $+,-,+$. Second, the three columns $X_0,X_1,X_2$ have no common support: every edge type is missing from at least one of them. The first property separates the columns $X_t$ from the stars, and the second prevents a single additional generator from serving all three of them.

Before we give the proof, we explain why it has to use the magnitudes of the entries.

\begin{remark}\label{rem:standard}
The column space of $S$ is spanned by $X_0$ and the columns of $R$, because $X_t=X_0+Rb_t$ for suitable vectors $b_t$ (see Section~\ref{sec:geometry}). Since $\operatorname{rank}R=11$, and since $D$ vanishes on the image of $R$ but not at $X_0$, the ordinary rank of $S$ is $12$. Moreover, the support of every column of $S$ is a union of stars: $\supp X_t$ is the union of the stars of the four nodes in $C_{2t+3}\cup C_{2t+4}$, and adding $LR_v$ adds $\st(v)$. Hence the twelve rectangles $\st(v)\times\{\text{columns whose support contains }\st(v)\}$ cover the support of $S$, and the rectangle covering bound for $\rk S$ is at most $12$. For $S\ot S$, the same rectangles in each of the two blocks cover the support, and the ordinary rank is at most $24$. Thus neither the rank nor any argument that only uses the supports of these matrices can prove the bounds $14$ and $27$.
\end{remark}

\begin{proposition}\label{prop:ranks}
We have $\rk S=14$ and $\rk(S\ot S)=27$.
\end{proposition}

\begin{proof}
The upper bounds were proved in Lemma~\ref{lem:upper}, and we prove both lower bounds at once. We fix $d\in\{1,2\}$ and let $S^{(d)}$ be $S$ if $d=1$ and $S\ot S$ if $d=2$. We regard $S^{(d)}$ as a matrix with $d$ row blocks, each indexed by the edges of a copy of $G$; its columns are the $d$-tuples of columns of $S$. Suppose, for a contradiction, that the columns of $S^{(d)}$ lie in the cone of a set $\mathcal G$ of $r\le13d$ nonzero nonnegative vectors, which we call \emph{generators}. We normalize every $g\in\mathcal G$ so that $\|g\|_\infty=1$. Then the coefficient of $g$ in any nonnegative representation of a vector $x$ is at most $\|x\|_\infty$, as one sees by evaluating at a coordinate where $g$ equals one. The proof proceeds through four claims.

\medskip\noindent\textbf{Claim 1.} \emph{A generator is assigned to at most two nodes, and if it is assigned to two nodes, they are adjacent and lie in the same block.}
For each of the $12d$ nodes $v$ of the $d$ copies of $G$, we consider the column of $S^{(d)}$ with $X_0+LR_v$ in the block of $v$ and $X_0$ in every other block, divide it by $L$, and call the result the \emph{probe} $Y_v$. Let $R_v^{(d)}$ be the vector equal to $R_v$ in the block of $v$ and zero elsewhere; every entry of $Y_v$ differs from the corresponding entry of $R_v^{(d)}$ by at most $L^{-1}$. We fix a representation $Y_v=\sum_g\lambda_{g,v}\,g$ with $\lambda_{g,v}\ge0$; then $\lambda_{g,v}\le1+L^{-1}<2$. We use the constants
\[
  \tau=\frac1{5000},\qquad\beta=\frac1{L\tau}=\frac1{200},\qquad
  \varepsilon=\frac1L+26\tau=\frac{5201}{10^6},
\]
and we say that $g$ is \emph{assigned} to $v$ if $\lambda_{g,v}\ge\tau$. Outside $\st(v)$ every entry of $Y_v$ is at most $L^{-1}$, so every entry of an assigned generator outside $\st(v)$ is at most $\beta$. Since $\beta<1=\|g\|_\infty$, a generator assigned to two nodes attains its maximum on the intersection of their stars. Hence the two nodes lie in the same block and are adjacent, the generator is at most $\beta$ outside their common edge, and we say that it is \emph{centered} at this edge. No generator is assigned to three nodes, because three stars have empty intersection.

\medskip\noindent\textbf{Claim 2.} \emph{Every node has a generator assigned to it alone.}
Suppose that $v$ has none, and let $e\in\st(v)$. If no generator assigned to $v$ is centered at $e$, then every generator assigned to $v$ contributes less than $2\beta$ to the entry $Y_v(e)$, and every other generator contributes less than $\tau$. This gives $Y_v(e)<26(\tau+2\beta)=663/2500<1$, which contradicts $Y_v(e)\ge1$. Hence each of the four edges of $\st(v)$ carries a generator centered there. Now let $b\ge1$ be the number of nodes without a private generator. The other $12d-b$ nodes have distinct private generators, and the $b$ nodes require further generators centered at the edges of their stars: four if $b=1$; at least seven if $b=2$, because two stars contain at least seven distinct edges; and at least $2b$ if $b\ge3$, because a centered generator serves at most two nodes. The total is at least $12d+3$, $12d+5$ or $12d+b$, respectively, and in each case it exceeds $13d$ for $d\le2$. This contradiction proves the claim. We reserve one private generator $g_v$ for every node $v$. At most $d$ further generators remain, and we call them \emph{extras}.

\medskip\noindent\textbf{Claim 3.} \emph{Every small target uses an unassigned extra with a positive coefficient.}
A \emph{small target} is a column $(X_{t_1},\dots,X_{t_d})$ of $S^{(d)}$. We call a node \emph{exceptional} if some extra is assigned to it. At a node $v$ that is not exceptional, $g_v$ is the only generator assigned to $v$. Writing $\lambda=\lambda_{g_v,v}$, the other generators together contribute less than $25\tau$ to every entry of $Y_v$, and therefore
\begin{equation}\label{eq:reserved}
  \|\lambda g_v-R_v^{(d)}\|_\infty\le\varepsilon,\qquad\lambda\ge1-\varepsilon .
\end{equation}
Since $D(R_v^{(d)})=0$ by~\eqref{eq:cycle}, every alternating sum $D$ on a single block satisfies $|D(g_v)|\le6\varepsilon/(1-\varepsilon)$.

There are at most $d$ extras, and each extra is assigned only to nodes of one block, so some block contains the nodes of at most one assigned extra. In this block we choose a six-cycle that avoids these nodes, which form a single node or the two ends of an edge; this is possible because every cluster contains a second node. On the edges of this cycle, every assigned extra and every reserved generator at an exceptional node has entries at most $\beta$. The same holds for the extras assigned in other blocks, which are at most $\beta$ on the entire block. The coefficients in a representation of a small target are at most one. Hence the reserved generators and the assigned extras together contribute at most
\[
  26\max\Bigl\{6\beta,\ \frac{6\varepsilon}{1-\varepsilon}\Bigr\}=\frac{62412}{76523}<1
\]
in absolute value to the alternating sum of the target, which equals one by~\eqref{eq:cycle}. Thus some unassigned extra has a positive coefficient. If no extra is assigned at all, there are no exceptional nodes, and the same bound holds on every block and for every cycle.

\medskip\noindent\textbf{Claim 4.} \emph{The unassigned extras cannot serve all small targets.}
A nonzero generator $g$ has a positive coordinate at some edge of some block, and some $X_t$ vanishes at this edge. Hence $g$ has coefficient zero in every small target whose entry in that block is $X_t$, and we say that $g$ is \emph{forbidden} at these targets. By Claim~3 every small target uses an unassigned extra, so there are at least two unassigned extras. This is impossible for $d=1$, and we obtain $\rk S\ge14$.

For $d=2$ there are exactly two extras, and both are unassigned. Let $T$ be a small target at which the first extra is forbidden. By the last sentence of Claim~3, on each block the reserved generators contribute less than one in absolute value to the alternating sum of $T$. Hence the second extra contributes on both blocks, and in particular it has a positive coordinate in each block. By symmetry, the same holds for the first extra. A positive coordinate in the first block forbids at least one of the three values of $t_1$, and a positive coordinate in the second block forbids at least one value of $t_2$. Thus each extra can be used in at most $2\cdot2=4$ of the nine small targets $(X_{t_1},X_{t_2})$, and the two extras together in at most eight. This contradicts Claim~3 and proves $\rk(S\ot S)\ge27$.
\end{proof}

The argument stops at $d=2$: for $d=3$ the count in Claim~2 no longer exceeds the budget $13d$. We do not know the nonnegative rank of the threefold $1$-product $S\ot S\ot S$.

\section{The polytope}\label{sec:geometry}

We now realize $S$ geometrically. Let $B\subset\R^{24}$ be the convex hull of the $39$ columns of $S$. We determine all facets and vertices of $B$ and show that the coordinate inequalities $x_e\ge0$ are among its facets. Wedging over the remaining facets then produces $P$.

\subsection{Coordinates}
The kernel of $R$ is spanned by the vector $\kappa\in\R^{V(G)}$ with entries $+1$ on the even clusters and $-1$ on the odd clusters. Indeed, $Ry=0$ means that $y_v=-y_w$ for every edge $vw$, and $G$ is connected and bipartite. Hence $\operatorname{rank}R=11$. Let $\iota_j$ be the indicator vector of $C_j$. A direct check shows that $X_t=X_0+Rb_t$ for
\[
  b_0=0,\qquad b_1=\iota_0-\iota_3,\qquad b_2=\iota_1-\iota_4 .
\]
We therefore write the points of the affine hull of $B$ as $x=X_0+Ry$ with \emph{node potentials} $y\in\R^{V(G)}$. The potentials are determined up to adding multiples of $\kappa$, and we fix the gauge by requiring $y_{v_0}=0$ for a fixed node $v_0\in C_5$. This identifies the affine hull of $B$ with $\R^{11}$. In these coordinates
\begin{equation}\label{eq:mink}
  B=\conv\{X_0,X_1,X_2\}+L\conv\bigl(\{0\}\cup\{R_v:v\in V(G)\}\bigr),
\end{equation}
because the columns of $S$ are exactly the sums of a point of the first set and a point of the second. In particular, $B$ is $11$-dimensional.

\subsection{Facets and vertices}
For a node $v$ in an even cluster and a node $w$ in an odd cluster we put $\gamma_{vw}=1$ if $vw$ is an edge of type $2$, $3$ or $4$, and $\gamma_{vw}=0$ otherwise. The pairs with $\gamma_{vw}=0$ include the twelve pairs of nonadjacent nodes, which join opposite clusters $C_j$ and $C_{j+3}$. Since $\gamma_{vw}$ depends only on the clusters $C_j\ni v$ and $C_k\ni w$, we also write $\gamma_{jk}$.

\begin{proposition}\label{prop:base}
The polytope $B$ is defined by the $37$ inequalities
\begin{equation}\label{eq:h}
  y_v+y_w+\gamma_{vw}\ge0\quad(v\text{ even},\ w\text{ odd}),\qquad\sum_{v\in V(G)}y_v\le L,
\end{equation}
each of which defines a facet, and its vertices are exactly the $39$ columns of $S$. For every edge $e=vw$, the inequality $y_v+y_w+\gamma_{vw}\ge0$ is the coordinate inequality $x_e\ge0$.
\end{proposition}

\begin{proof}
For an edge $e=vw$ we have $x_e=(X_0)_e+y_v+y_w$ and $(X_0)_e=\gamma_{vw}$, which proves the last assertion. We next derive~\eqref{eq:h}. By~\eqref{eq:mink}, the point $y$ lies in $B$ if and only if there are $\alpha_1,\alpha_2\ge0$ with $\alpha_1+\alpha_2\le1$ and $a\in\R$ such that $u=y-\alpha_1b_1-\alpha_2b_2+a\kappa$ is nonnegative with coordinate sum at most $L$. Since $\kappa$, $b_1$ and $b_2$ have coordinate sum zero, the last condition reads $\sum_vy_v\le L$. We put $m_j=\min_{v\in C_j}y_v$. For fixed $a$, the nonnegativity of $u$ on the six clusters reads
\[
  a-m_3\le\alpha_1\le m_0+a,\qquad-m_4-a\le\alpha_2\le m_1-a,\qquad-m_2\le a\le m_5 .
\]
Taking $\alpha_1$ and $\alpha_2$ as small as these conditions allow, we see that admissible values with $\alpha_1+\alpha_2\le1$ exist if and only if
\[
\begin{gathered}
  m_0+m_3\ge0,\qquad m_1+m_4\ge0,\qquad m_3+m_4\ge-1,\\
  \max\{-m_2,-m_0,-m_4-1\}\le a\le\min\{m_5,m_1,m_3+1\}.
\end{gathered}
\]
We eliminate $a$ by comparing each lower bound with each upper bound, and we discard the three comparisons that are implied by the first three inequalities. What remains are nine inequalities $m_j+m_k\ge-\gamma_{jk}$, one for each even cluster $j$ and each odd cluster $k$, with the right-hand sides listed in Table~\ref{tab:cluster}. Since $m_j$ is a minimum over $C_j$, each of them is equivalent to the four inequalities in~\eqref{eq:h} for the pairs of nodes in $C_j\times C_k$.

\begin{table}[ht]
\centering
\begin{tabular}{c|ccc}
 $-\gamma_{jk}$ & $k=1$ & $k=3$ & $k=5$\\ \hline
 $j=0$ & $0$ & $0$ & $0$\\
 $j=2$ & $0$ & $-1$ & $0$\\
 $j=4$ & $0$ & $-1$ & $-1$
\end{tabular}
\caption{The lower bounds for $m_j+m_k$ in the description of $B$. Each entry stands for four inequalities in~\eqref{eq:h}, one for each pair of twins. The entries $-1$ correspond to the edge types $2$, $3$ and $4$.}
\label{tab:cluster}
\end{table}

Each inequality in~\eqref{eq:h} defines a facet. For the pair $(v,w)$ we set $y_v=0$, $y_w=-\gamma_{vw}$ and $y_u=2$ for all other nodes $u$; for the sum inequality we set all potentials equal to $L/12$. After a shift by a multiple of $\kappa$ that restores the gauge, which changes no inequality, we obtain in each case a point of $B$ at which exactly one of the $37$ inequalities is tight.

Finally, we show that the $39$ columns are vertices. At $X_0$, where $y=0$, the tight inequalities are those with $\gamma_{vw}=0$: the edges of the types $5$, $0$ and $1$, and the pairs between opposite clusters. These cluster pairs form a connected graph on the six clusters, and each of them contributes all four node pairs, so the tight pairs form a connected bipartite graph on $V(G)$. Along such a graph the equations $y_v+y_w=-\gamma_{vw}$ determine $y$ up to a multiple of $\kappa$, which the gauge removes; hence $X_0$ is a vertex. The same argument applies to $X_1$ and $X_2$ after shifting the cluster indices by two, because each $b_t$ gives opposite potentials to the two ends of every opposite pair, so all opposite-pair inequalities remain tight. At $X_t+LR_v$ the tight pairs not involving $v$ are the same as at $X_t$, and they still connect the eleven nodes other than $v$, since every cluster retains a node. These equations determine the eleven potentials up to a multiple of $\kappa$, and the sum inequality $\sum_u y_u=L$ is tight. Here $\kappa$ sums to zero on $V(G)$ and has $\kappa_{v_0}=-1$ because $v_0\in C_5$. If $v\neq v_0$, the gauge $y_{v_0}=0$ fixes the multiple of $\kappa$ on $V(G)\setminus\{v\}$, after which the sum equation determines $y_v$; if $v=v_0$, the gauge sets $y_{v_0}=0$ and the sum of $\kappa$ on the remaining eleven nodes equals $-\kappa_{v_0}=1\neq0$, so the sum equation fixes the multiple of $\kappa$. Hence all $39$ columns are vertices of $B$.
\end{proof}

\subsection{The polytope \texorpdfstring{$P$}{P} and the proof of Theorem~\ref{thm:main}}
We retain the $24$ coordinate inequalities of $B$ and wedge over its other $13$ facets, namely the twelve inequalities for pairs of opposite clusters and the sum inequality. In the notation of~\eqref{eq:wedge}, $P=\widehat B$; explicitly,
\begin{equation}\label{eq:P}
P=\left\{(y,z)\in\R^{V(G)}\times\R^{13}\ :\
\begin{array}{ll}
 y_{v_0}=0, & \\[2pt]
 y_v+y_w+\gamma_{vw}\ge0 & (vw\in E(G)),\\[2pt]
 0\le z_{vw}\le y_v+y_w & (v\in C_j,\ w\in C_{j+3},\ j=0,1,2),\\[2pt]
 0\le z_0\le L-\sum_{v}y_v &
\end{array}\right\}.
\end{equation}
Thus $P$ is defined by $50$ integer inequalities in $24$ free variables.

The vertices of $P$ follow from Step~1 of the proof of Theorem~\ref{thm:transfer}. At $X_t$ the only positive slack among the $13$ wedged facets is that of the sum inequality, since $b_t$ gives opposite potentials to the ends of every opposite pair. At $X_t+LR_v$ the sum inequality is tight, and the only positive slacks are those of the two opposite pairs containing $v$. Hence $P$ has $3\cdot2+36\cdot4=150$ vertices. Over an interior point of $B$ all $13$ intervals are nondegenerate, so $\dim P=11+13=24$, and $P$ has $24+2\cdot13=50$ facets.

\begin{proof}[Proof of Theorem~\ref{thm:main}]
By Proposition~\ref{prop:base}, the $24$ retained facet rows of the slack matrix of $B$, evaluated at the $39$ vertices, form exactly the matrix $S$, since the slack of $x_e\ge0$ at a vertex $x$ is $x_e$. Theorem~\ref{thm:transfer} with $k=13$ and Proposition~\ref{prop:ranks} give
\[
  \xc(P)=\rk S+26=40\qquad\text{and}\qquad\xc(P\times P)=\rk(S\ot S)+52=79.\qedhere
\]
\end{proof}

\subsection{An explicit extension of the square}
The upper bound $\xc(P\times P)\le79$ can also be seen directly, without Theorem~\ref{thm:transfer}. We write $x=X_0+Ry$ and $x'=X_0+Ry'$ for the two factors, keep the $26$ wedge inequalities of each factor, introduce $h\in\R^{27}_{\ge0}$ and impose $(x,x')=Wh$, where the columns of $W$ are the $27$ vectors in~\eqref{eq:productdict}. Since $W\ge0$, the equation forces $x\ge0$ and $x'\ge0$, which are the $48$ retained inequalities. Conversely, Lemma~\ref{lem:upper} represents every pair of columns of $S$ in the form $Wh$ with $h\ge0$, and hence every pair of points of $B$, which are convex combinations of columns. Thus the system projects onto $P\times P$. Each column of $W$ is nonnegative and nonzero, so $h$ is bounded whenever $x$ and $x'$ are. The resulting extension has at most $2\cdot26+27=79$ facets.

\input{binary_deficit}

\section{Discussion}\label{sec:discussion}

\subsection{Consequences}
We first record two consequences of Theorem~\ref{thm:main}.

\begin{corollary}\label{cor:copies}
For every $m\ge1$,
\[
  \xc(P^{2m})\le79m=2m\,\xc(P)-m .
\]
Moreover, the limit $\lim_{n\to\infty}\xc(P^n)/n$ exists and is at most $79/2$.
\end{corollary}

\begin{proof}
Taking $m$ copies of an optimal extension of $P\times P$ gives the inequality. The sequence $n\mapsto\xc(P^n)$ is subadditive, so by Fekete's lemma $\xc(P^n)/n$ converges to its infimum, which is at most $\xc(P^2)/2=79/2$.
\end{proof}

Thus the number of inequalities saved, relative to modeling the $2m$ copies separately, grows at least linearly in $m$. The next corollary shows that the phenomenon is not tied to dimension $24$, and that it disappears under the operation for which Tiwary, Weltge and Zenklusen proved additivity.

\begin{corollary}\label{cor:dimensions}
For every $k\ge1$, the $(24+k)$-dimensional polytope $P\times\Delta_k$ satisfies
\[
  \xc\bigl((P\times\Delta_k)^2\bigr)=81+2k<82+2k=2\,\xc(P\times\Delta_k).
\]
On the other hand, the pyramid $\pyr(P)$ over $P$ satisfies $\xc(\pyr(P)\times\pyr(P))=2\,\xc(\pyr(P))$.
\end{corollary}

\begin{proof}
Grande, Padrol and Sanyal proved that $\xc(Q\times\Delta_k)=\xc(Q)+k+1$ for every polytope $Q$ \cite[Corollary~2.4]{gps}. Applied once, this gives $\xc(P\times\Delta_k)=41+k$. Since $(P\times\Delta_k)^2$ is affinely isomorphic to $(P\times P)\times\Delta_k\times\Delta_k$, applying it twice gives $\xc((P\times\Delta_k)^2)=79+2(k+1)$. The second statement is a special case of the theorem of Tiwary, Weltge and Zenklusen \cite{twz}.
\end{proof}

Together with Theorem~\ref{thm:main} in dimension $24$, this gives strictly subadditive squares in every dimension at least $24$.

\paragraph{Free sums and joins.}
Translate full-dimensional realizations of $U,V$ so that each contains the origin in its interior in its own space. Their free sum is $U\oplus V=\conv((U,0)\cup(0,V))$. Its polar is $U^\circ\times V^\circ$, since a linear functional is at most one on the free sum precisely when its two restrictions are at most one on the respective factors. Polarity preserves extension complexity: in the facet description normalized to right-hand side one, the slack matrix of the polar is the transpose of that of the original polytope, up to order. Hence $\xc(U^\circ)+\xc(V^\circ)-\xc(U^\circ\oplus V^\circ)=\delta(U,V)$, and by Theorem~\ref{thm:binary} the free sum of two copies of the same polytope can also have an arbitrarily large deficit. For the join, place the factors in affinely independent flats. Each facet of the join is the join of a facet of one factor with the whole other factor; extending the corresponding affine slack by zero on that other flat gives a block-diagonal slack matrix. Its nonnegative rank is the sum of the block ranks, so extension complexity is additive for joins of positive-dimensional factors.

\paragraph{An upper bound for the deficit.}
The face bound of Grande, Padrol and Sanyal \cite[Corollary~2.2]{gps} gives $\xc(D)\ge\xc(F)+\dim D-\dim F$ for a nonempty face $F$ of $D$. Applying it to $U\times\{v\}$ at a vertex of $V$ gives $\xc(U\times V)\ge\xc(U)+\dim V$. Interchanging the factors yields
\[
  0\le \delta(U,V)\le\min\{\xc(U)-\dim U,\ \xc(V)-\dim V\}.
\]
Thus an unbounded deficit requires unbounded excess over dimension on both sides. For the polytope $P$ of Theorem~\ref{thm:main} this bound equals $\xc(P)-\dim P=16$, while the actual deficit is one.

\subsection{Zero-error channel simulation}\label{sec:interpretations}
A sender sees an input $u$ of a channel $W$, given by a column-stochastic matrix, and transmits one message without shared randomness; the receiver must output an exact sample of $W(\,\cdot\,|\,u)$. The least number of messages is $\rk W$, because a column-stochastic matrix has a nonnegative factorization of minimum size with column-stochastic factors \cite[Theorem~3.2]{cohenrothblum}, which serve as encoder and decoder. By Minkowski's relation for facet normals, a positive rescaling of the rows turns $S_Q$ into a column-stochastic matrix, so $\xc(Q)$ is the least alphabet size for this \emph{slack channel} of $Q$. The $1$-product corresponds to a labeled mixture: the sender sees one input of each of two slack channels, and the receiver outputs a fair random label together with a sample of the corresponding channel. This channel is a rescaled slack matrix of $Q\times Q'$, so it needs exactly $\xc(Q\times Q')$ messages. If every message must determine the label, then $\xc(Q)+\xc(Q')$ messages are necessary, since after fixing the second input the messages with the first label simulate the first channel, and they clearly suffice. For $Q=P$ and $Q'=P\times\Delta_{48}$, Theorem~\ref{thm:main} and the simplex theorem of \cite{gps} give $40+89=129$ messages with determined labels, against $\xc(P\times P\times\Delta_{48})=128=2^7$. Sharing messages between the two labels thus shortens a fixed-length message from $8$ bits to $7$.

\subsection{Open questions}

Is there a sequence of polytopes $R_i$ with $\xc(R_i)\to\infty$ for which the ratios $\delta(R_i,R_i)/(2\xc(R_i))$ are bounded below by a positive constant? Theorem~\ref{thm:binary} gives a uniform relative bound for different factors. For identical factors it gives an unbounded absolute saving. Other structural restrictions lead to their own questions. How small can the dimension of a strictly subadditive square be? The example of \cite[Section~6.1]{shitovoverview} has dimension $59$, and Theorem~\ref{thm:main} with Corollary~\ref{cor:dimensions} gives every dimension at least $24$. Does additivity hold for $0/1$ polytopes, or for matroid base polytopes under direct sum? The latter restriction was part of the original motivation \cite{gps}. Also, what is the infimum of $\xc(U\times V)/(\xc(U)+\xc(V))$ over positive-dimensional factors? The face bound gives a lower bound of $1/2$, and Theorem~\ref{thm:main} gives an upper bound of $79/80$.

For a fixed polytope define the asymptotic cost $\xc^\infty(U)=\lim_{m\to\infty}\xc(U^m)/m$. The limit exists by subadditivity, and Corollary~\ref{cor:copies} gives $\xc^\infty(P)\le79/2<\xc(P)$. Determining whether $\xc^\infty(P)<79/2$ would quantify further gains from grouping copies of this fixed example.

\section*{Verification and ancillary files}
All lower bounds are proved in the text, for arbitrary real nonnegative factorizations. The scripts described here check the finite inputs of these proofs. The verification code and exact data are available at \repositoryURL, and the accompanying README gives the commands. For the exact square, the SymPy script \texttt{verify\_construction.py} reconstructs the base, the two dictionaries and the alternating sums. The standard-library script \texttt{check\_lower\_bounds.py} checks the exceptional-node patterns, the rational bounds in Proposition~\ref{prop:ranks}, and the star cover of Remark~\ref{rem:standard}. The cddlib program \texttt{independent\_check.py} recovers the $150$ vertices from the $50$ halfspaces in exact arithmetic. The script \texttt{check\_full\_slack.py} checks all $2{,}250{,}000$ entries of the square slack matrix by integer multiplication.

For Section~\ref{sec:binary}, \texttt{verify\_family.py} checks incidence ranks, label choices and the conditional counts. The program \texttt{verify\_counting.py} independently exhausts the four-block label profiles. A separate FLINT implementation, \texttt{verify\_unbounded\_note.py}, directly checks the rank after deletion for each of the $1{,}080$ and $1{,}728$ retained coordinate facets. For the square construction, \texttt{check\_identical\_factors.py} enumerates all five-profile candidates for the first two alphabets, checks a six-profile cover, and checks the ten shared twin profiles on all $11{,}025$ paired inputs. The general bounds follow from the transfer theorem, Proposition~\ref{prop:familycost} and Lemmas~\ref{lem:largealphabet}--\ref{lem:profilebridge}.

\paragraph{Acknowledgment.}
I thank Yaroslav Shitov for informing me of the earlier counterexample \cite{shitovoverview}.

\paragraph{Use of AI tools.}
During this work the author used the AI systems OpenAI Codex (Astra~6.2), Anthropic Claude and Google Antigravity (Carbon~4). They were used to explore candidate constructions and to test proof strategies, to write and run the verification code, and to edit the text. The author selected and corrected these suggestions, completed the proofs, and checked every statement and proof in the paper, and takes full responsibility for its content.

\end{document}

%% file: binary_deficit.tex
\section{Unbounded savings in a binary product}\label{sec:binary}

We first prove the assertions about $U_n,V_n$ in Theorem~\ref{thm:binary}. We construct two matrices, each with $n$ independent coordinates. The upper bound for their $1$-product is immediate, because a two-element subset and a three-element subset of a four-element set always intersect. The main work lies in the lower bounds for the two individual matrices, where a factorization may share generators among all $n$ coordinates.

\subsection{The two dictionaries and their joint use}\label{sec:dict}
Write $[4]=\{1,2,3,4\}$ and $\mathcal T_k=\binom{[4]}{k}$ for $k=2,3$. Let $\Gamma_k$ be the bipartite incidence graph with left nodes $s\in[4]$, right nodes $T\in\mathcal T_k$, and an edge $(s,T)$ exactly when $s\in T$. Both graphs have twelve edges and maximum degree three. For a node $f$ let $I(f)$ be the set of its incident edges. We call the nodes and edges of $\Gamma_k$ \emph{coarse}.

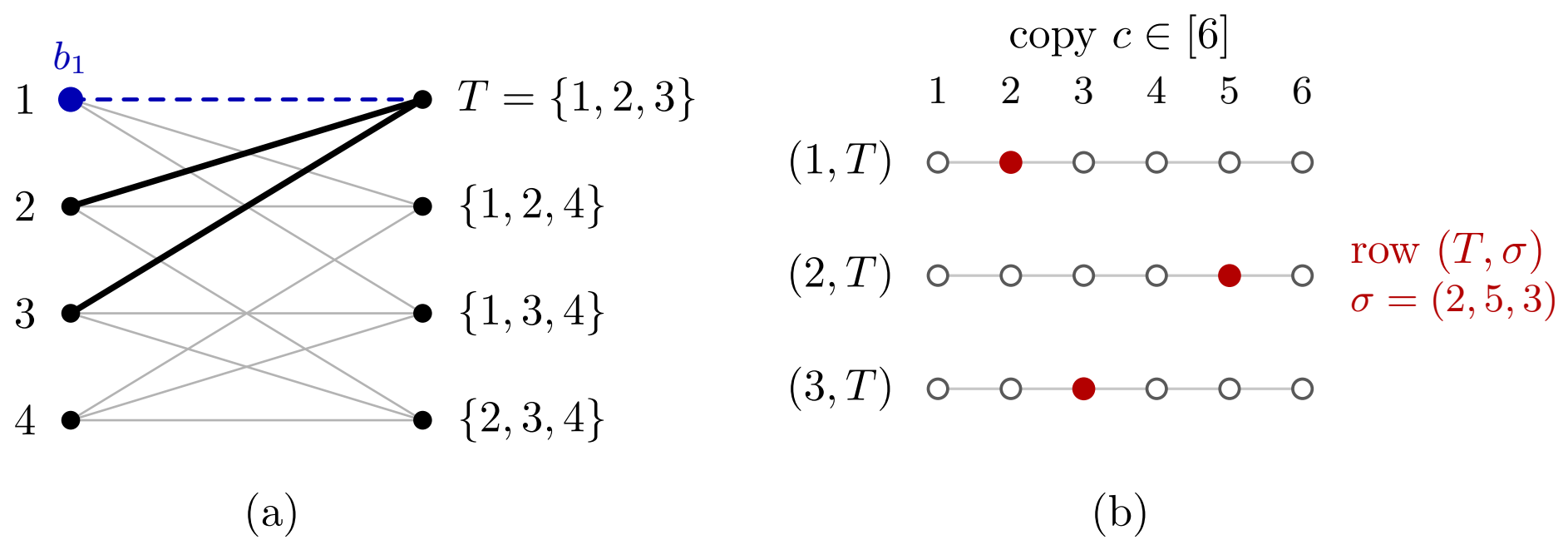
\begin{figure}[htb]
\centering
\begin{tikzpicture}[line join=round,line cap=round,
    vtx/.style={circle,fill=black,inner sep=1.5pt},
    svtx/.style={circle,fill=blue!70!black,inner sep=2pt}]
  \begin{scope}
    \foreach \s/\y in {1/2.55,2/1.70,3/0.85,4/0.0}{
      \coordinate (L\s) at (0,\y);
      \node[left=4pt,font=\small] at (L\s) {$\s$};
    }
    \coordinate (R1) at (2.8,2.55); \node[right=4pt,font=\small] at (R1) {$T=\{1,2,3\}$};
    \coordinate (R2) at (2.8,1.70); \node[right=4pt,font=\small] at (R2) {$\{1,2,4\}$};
    \coordinate (R3) at (2.8,0.85); \node[right=4pt,font=\small] at (R3) {$\{1,3,4\}$};
    \coordinate (R4) at (2.8,0.0);  \node[right=4pt,font=\small] at (R4) {$\{2,3,4\}$};
    \foreach \s/\r in {1/2,2/2,4/2, 1/3,3/3,4/3, 2/4,3/4,4/4}{
      \draw[black!30,line width=0.5pt] (L\s)--(R\r);
    }
    \draw[dashed,blue!70!black,line width=0.9pt] (L1)--(R1);
    \draw[black,line width=1.4pt] (L2)--(R1);
    \draw[black,line width=1.4pt] (L3)--(R1);
    \foreach \s in {2,3,4}{\node[vtx] at (L\s){};}
    \node[svtx] at (L1){};
    \node[above=2pt,blue!70!black,font=\footnotesize] at (L1) {$b_1$};
    \foreach \r in {1,2,3,4}{\node[vtx] at (R\r){};}
    \node[font=\small] at (1.6,-0.75) {(a)};
  \end{scope}
  \begin{scope}[xshift=6.9cm]
    \foreach \c in {1,...,6}{
      \pgfmathsetmacro{\xc}{0.58*(\c-1)}
      \node[font=\footnotesize] at (\xc,2.62) {$\c$};
    }
    \node[font=\small] at (1.45,3.05) {copy $c\in[6]$};
    \foreach \idx/\lbl/\yr/\pick in {1/{$(1,T)$}/2.05/2, 2/{$(2,T)$}/1.15/5, 3/{$(3,T)$}/0.25/3}{
      \node[left=6pt,font=\small] at (0,\yr) {\lbl};
      \draw[black!22,line width=0.6pt] (0,\yr)--(2.9,\yr);
      \foreach \c in {1,...,6}{
        \pgfmathsetmacro{\xc}{0.58*(\c-1)}
        \ifnum\c=\pick
          \fill[red!70!black] (\xc,\yr) circle (2.6pt);
        \else
          \fill[white] (\xc,\yr) circle (2.2pt);
          \draw[black!65,line width=0.7pt] (\xc,\yr) circle (2.2pt);
        \fi
      }
    }
    \node[right=7pt,red!70!black,font=\small,align=left] at (2.9,1.15) {row $(T,\sigma)$\\[-1pt]\footnotesize $\sigma=(2,5,3)$};
    \node[font=\small] at (1.45,-0.75) {(b)};
  \end{scope}
\end{tikzpicture}
\caption{The block dictionary for $k=3$. (a) The bipartite incidence graph $\Gamma_3$. For $T=\{1,2,3\}$ and the choice $s=1$ in~\eqref{eq:familychoice}, the vector $b_1$ supplies the lifted coordinate at $1$, while summing all $N=6$ star copies along each of the two heavy edges $(2,T)$ and $(3,T)$ supplies $2$ and $3$ once and $T$ twice. (b) The $18$ stars incident to $T$, arranged by incident coarse edge and copy index $c\in[6]$. A single row $(T,\sigma)$ selects one copy on each of the three incident edges (solid red); fixing $\sigma(e)=c$ on one edge leaves all $6$ choices equally frequent on any other edge.}
\label{fig:choice}
\end{figure}

Fix $N=6$. For each coarse edge $e$ we introduce $N$ copies $(e,c)$, $c\in[N]$, and for each coarse node $f$ we introduce one row $(f,\sigma)$ for every choice function $\sigma\colon I(f)\to[N]$. The copy $(e,c)$ defines a zero-one vector $g_{e,c}$ whose entry at the row $(f,\sigma)$ is one precisely when $e\in I(f)$ and $\sigma(e)=c$. We call these $p=12N=72$ vectors \emph{stars}; see Figure~\ref{fig:choice}. Thus a row meets exactly one copy of each incident edge, and hence at most three stars. A vector of values on the coarse nodes \emph{lifts} to the new row set by repeating its value at $f$ at every row $(f,\sigma)$.

We suppress $k$ in the notation for individual vectors when the graph is fixed. Let $b_s$ be the lifted indicator of the left node $s$. For $T\in\mathcal T_k$, let $v_T$ be the lift of the vector with value one at each left node in $T$, value $k-1$ at the right node $T$, and value zero elsewhere. For each $s\in T$ we have the representation
\begin{equation}\label{eq:familychoice}
 v_T=b_s+\sum_{t\in T\setminus\{s\}}\sum_{c=1}^N g_{(t,T),c}.
\end{equation}
Indeed, summing the $N$ copies of an edge gives the lifted indicator of its two endpoints. Hence the double sum supplies each of the $k-1$ remaining left nodes of $T$ once and the right node $T$ exactly $k-1$ times.

Let $\Lambda$ be a positive integer, and let $W_k(\Lambda)$ be the matrix whose columns are the vectors $v_T$ and $v_T+\Lambda g_{e,c}$ for all $T\in\mathcal T_k$ and all stars $g_{e,c}$. We write $M_{k,n}=W_k(\Lambda)^{[n]}$ for the iterated $1$-product of $n$ copies of $W_k(\Lambda)$. Its columns are the stacked $n$-tuples of independently chosen columns of $W_k(\Lambda)$, and we call the group of rows belonging to one coordinate a \emph{block}. In each block, a column with index $T$ can be written with any label $s\in T$ as in~\eqref{eq:familychoice}, because a star shift $\Lambda g_{e,c}$ is supplied by the star itself. Three fixed labels meet every two-element subset of $[4]$, and two fixed labels meet every three-element subset. Using the $p$ stars and these labels separately in each block gives
\begin{equation}\label{eq:familyupper}
  \rk M_{2,n}\le (p+3)n,\qquad \rk M_{3,n}\le(p+2)n .
\end{equation}
Now we pair the $i$th block of $M_{2,n}$ with the $i$th block of $M_{3,n}$. For any pair of columns, the corresponding sets $T\in\mathcal T_2$ and $U\in\mathcal T_3$ intersect, because $|T|+|U|=5>4$, and we choose $s\in T\cap U$. Both blocks can then use the label $s$ through a single generator, namely the vector that equals $b_s$ in each of the two blocks, where $b_s$ is formed in the respective graph. There are four such vectors for each pair of blocks. Together with the $2p$ one-sided stars per pair, this proves
\begin{equation}\label{eq:familyjoint}
             \rk(M_{2,n}\ot M_{3,n})\le(2p+4)n .
\end{equation}

\paragraph{Label profiles.}
Both lower bounds in this section record which labels a generator requires. We set this up for a general list of blocks, in which block $i$ has a finite label set $L_i$ and a family $\mathcal A_i$ of subsets of $L_i$. An \emph{input} is a tuple $(T_1,\ldots,T_n)$ with $T_i\in\mathcal A_i$. A \emph{profile} consists of a nonempty set $J$ of blocks, which it is said to \emph{specify}, and a label $\ell_i\in L_i$ for each $i\in J$; it is \emph{available} at an input if $\ell_i\in T_i$ for every $i\in J$. A family of profiles \emph{covers} a set $I$ of blocks if, for every input and every $i\in I$, some available profile specifies $i$. The \emph{cover number} of the list is the least size of a family covering all blocks. A hitting set of $\mathcal A_i$, used as singleton-block profiles, covers block $i$; for $\mathcal T_2$ and $\mathcal T_3$ the least hitting sets have three and two elements.

In $M_{k,n}$ every block has $L_i=[4]$ and $\mathcal A_i=\mathcal T_k$, and we call the columns $(v_{T_1},\ldots,v_{T_n})$ \emph{small targets}. We label the rows: a row over a left node $s$ receives the label $s$, and a row over a right node $U$ receives a fixed element of $U$. If $v_T$ is positive at a row, then the label of that row belongs to $T$. For a nonzero nonnegative vector $z$ on the rows of $M_{k,n}$, let $J_z$ be the set of blocks in which $z$ has a positive entry. We choose one such row in each block of $J_z$ once and for all; the \emph{profile} of $z$ consists of $J_z$ together with the labels of these rows. If $z$ has a positive coefficient in a nonnegative representation of a small target $(v_{T_1},\ldots,v_{T_n})$, then every positive entry of $z$ lies in the support of the target, so the profile of $z$ is available at $(T_1,\ldots,T_n)$.

\subsection{The costs of the two separately optimized factors}
\begin{proposition}\label{prop:familycost}
Let $H=(p+3)n$ and $\Lambda=10^5H^3$. For this choice of scale,
\[
 \rk M_{2,n}=(p+3)n,\qquad
 \rk M_{3,n}\ge(p+32/27)n .
\]
\end{proposition}

\begin{proof}
Fix $k\in\{2,3\}$ and consider a nonnegative factorization of $M_{k,n}$ with $r\le H$ generators. We delete unused generators and normalize each remaining generator $w$ so that $\|w\|_\infty=1$. Then the coefficient of $w$ in a representation of a nonnegative vector is at most the infinity norm of that vector, as one sees at a coordinate where $w$ equals one. We regard the $pn$ stars of the $n$ blocks as vectors on the full row set of $M_{k,n}$, extending each by zero outside its own block.

\paragraph{Assigning generators to the star probes.}
Fix $T_0\in\mathcal T_k$. For each star $g$ we consider the column of $M_{k,n}$ with $v_{T_0}+\Lambda g$ in the block of $g$ and $v_{T_0}$ in every other block. Dividing this column and a fixed representation of it by $\Lambda$ gives the \emph{probe} $y_g$. We have $\|y_g-g\|_\infty\le2/\Lambda$, and every coefficient in the representation of a probe is less than two. We use the thresholds
\[
 \tau=\frac1{100H^2},\qquad \beta=\frac{2}{\Lambda\tau}=\frac1{500H},
 \qquad \epsilon=\frac2\Lambda+H\tau\le\frac1{90H},
\]
and say that a generator $w$ is \emph{assigned} to $g$ if its coefficient in $y_g$ is at least $\tau$. At a row outside the support of $g$ the probe is at most $2/\Lambda$, so an assigned generator is at most $\beta$ there. Since $\beta<1$, a row where $w$ attains its maximum lies in every star to which $w$ is assigned. In particular, all these stars belong to one block. At that row the choice function selects a single copy of each incident coarse edge, so the assigned stars come from distinct coarse edges, and there are at most three of them.

\paragraph{Paying for stars without an exclusive generator.}
We call a star \emph{deficient} if no generator is assigned to it alone, and we write $b$ for the number of deficient stars. For every other star we reserve one generator that is assigned to it alone; these $pn-b$ reserved generators are distinct. Let $a$ be the number of remaining generators that are assigned to at least one star, and let $u$ be the number of generators assigned to no star. Then
\begin{equation}\label{eq:familybudget}
                  r=pn-b+a+u .
\end{equation}
Consider a deficient star $g_{e,c}$ and fix an endpoint $f$ of $e$. In the block of this star we select the rows $(f,\sigma)$ with $\sigma(e)=c$, all of which belong to the star. Every generator assigned to $g_{e,c}$ is also assigned to a second star, and we choose one such second star for each of them. As shown above, the second star lies in the same block and on a different coarse edge $e'$. If $e'\notin I(f)$, it contains none of the selected rows. If $e'\in I(f)$, it contains the selected rows with one particular value of $\sigma(e')$; since each of the $N$ values of $\sigma(e')$ occurs equally often among the selected rows, it contains exactly a fraction $1/N$ of them.

If fewer than $N$ generators were assigned to $g_{e,c}$, some selected row would lie in none of the chosen second stars. At such a row each generator assigned to $g_{e,c}$ contributes at most $2\beta$ to the probe, and every other generator contributes less than $\tau$. The total is at most $H(2\beta+\tau)<1$, although the probe is at least one at this row. Hence at least $N$ generators are assigned to each deficient star, and each of them is assigned to at least two stars. These generators are among the $a$ remaining assigned generators, and each of them is assigned to at most three stars. Counting incidences gives
\begin{equation}\label{eq:familydeficient} b\le 3a/N. \end{equation}

\paragraph{A linear functional for each unaffected block.}
We \emph{mark} every star to which one of the $a$ remaining assigned generators is assigned, and we call a block \emph{good} if each of its coarse edges has at least one unmarked copy. A block that is not good has a coarse edge all of whose $N$ copies are marked. A generator is assigned to at most one copy of a given edge, so this takes $N$ distinct generators among the $a$, all assigned within that block. Since no generator is assigned to stars in two different blocks, there are at least $n-a/N$ good blocks.

Fix a good block $i$ and choose an unmarked copy $\gamma_e$ of each of its coarse edges $e$. At every coarse node $f$ of this block we select the row $(f,\gamma|_{I(f)})$. Let $\Phi_i$ be the linear functional that adds the entries at the selected rows of left nodes and subtracts the entries at the selected rows of right nodes; it involves at most ten entries, each with coefficient $1$ or $-1$, all in block $i$. The functional vanishes on every star: the chosen copy of an edge equals one at the selected rows of both endpoints, every other copy in block $i$ is zero at all selected rows, and stars in other blocks vanish on block $i$. At a small target $x=(v_{T_1},\ldots,v_{T_n})$, on the other hand, the selected rows carry $k$ ones on the left and a single entry $k-1$ on the right, so $\Phi_i(x)=1$.

We now bound the contributions of the assigned generators to $\Phi_i(x)$. An unmarked star $g$ is not deficient, and its reserved generator $w$ is the only generator assigned to it. All other coefficients in the probe $y_g$ are less than $\tau$, so the coefficient $c$ of $w$ satisfies $\|cw-g\|_\infty\le\epsilon$ and $c\ge1-\epsilon$, and therefore $|\Phi_i(w)|\le10\epsilon/(1-\epsilon)$. Every other assigned generator, that is, a reserved generator of a marked star or one of the $a$ remaining assigned generators, is assigned to some marked star. No row selected for $\Phi_i$ lies in that star, because the star either belongs to another block or is a copy that $\gamma$ avoids. Hence such a generator is at most $\beta$ on all selected rows. Every coefficient in a representation of $x$ is at most two. In total, the generators other than the $u$ unassigned ones contribute to $\Phi_i(x)$ a quantity of absolute value at most
\[
        2H\max\{10\beta,10\epsilon/(1-\epsilon)\}\le20/89<1 .
\]
Since $\Phi_i(x)=1$, it follows that for every small target and every good block $i$, some unassigned generator with a positive coefficient has a positive entry in block $i$.

\paragraph{Conditional counting.}
Consider the profiles of the $u$ unassigned generators, as defined in Section~\ref{sec:dict}. We have just seen that at every small target and for every good block $i$, some unassigned generator with a positive coefficient has a positive entry in block $i$. Its profile specifies $i$ and is available at the input of the target. Thus these $u$ profiles cover the good blocks.

Fix a good block $i$ and sets $T_j\in\mathcal T_k$ for all $j\ne i$. Among the unassigned generators whose profile specifies $i$, we call a generator a \emph{candidate} if its labels in the other blocks that it specifies belong to the respective sets $T_j$. Suppose that there are at most $4-k$ candidates. Their labels in block $i$ form a set of at most $4-k$ elements, so some $k$-set $T_i$ avoids all of them. At the resulting input no available profile specifies $i$, which contradicts the covering property. Hence there are at least $5-k$ candidates for every choice of the sets outside block $i$.

Now choose the sets $T_j$, $j\ne i$, independently and uniformly from $\mathcal T_k$. A fixed label belongs to a random $k$-set with probability $\rho=k/4$, so a generator $z$ whose profile specifies $i$ is a candidate with probability $\rho^{|J_z|-1}$. Taking expectations and summing over the good blocks gives
\[
    (5-k)\#\{\text{good blocks}\}
       \le \sum_z |J_z|\rho^{|J_z|-1}
       \le u\max_{d\ge1} d\rho^{d-1} ,
\]
where the sum is over the unassigned generators $z$. For $\rho=1/2$ the maximum over the integers $d\ge1$ is one. For $\rho=3/4$ it is $27/16$, attained at $d=3$ and $d=4$, as one sees by comparing successive terms. Hence $u\ge c_k(n-a/N)$ with $c_2=3$ and $c_3=32/27$. Substituting this bound and~\eqref{eq:familydeficient} into~\eqref{eq:familybudget} yields
\[
       r\ge(p+c_k)n+a\left(1-\frac{3+c_k}{N}\right)\ge(p+c_k)n
\]
because $N=6$. A factorization with more than $H$ generators satisfies both lower bounds trivially. Together with~\eqref{eq:familyupper} this proves the proposition.
\end{proof}

The constant $32/27$ cannot be improved by this counting argument alone. Take three blocks and $k=3$, and consider the fractional relaxation in which each profile carries a nonnegative weight. Give weight $1/18$ to each of the $4^3$ profiles that specify one label in every block. For any block $i$ and any fixed sets in the other two blocks, exactly $4\cdot3^2=36$ of these profiles are compatible with the fixed sets, so the candidates have total weight two, as the argument requires. The total weight is $64/18=32/9=3\cdot(32/27)$, which equals the lower bound for this relaxation.

For example, for $n=6$ we obtain $\rk M_{2,6}=450$ and $\rk M_{3,6}\ge440$, and the second bound is also attained. Indeed, any three $3$-element subsets of $[4]$ have a common label, so for each group of three coordinates four generators suffice in addition to the stars, one for each label, and two groups give $6p+8=440$. The joint upper bound~\eqref{eq:familyjoint} is $888$, so the saving at $n=6$ is at least two.

\subsection{Realization of the two-family construction}
Let $C_k(\Lambda)$ be the convex hull of the columns of $W_k(\Lambda)$. Before we apply Theorem~\ref{thm:transfer}, we verify that the coordinate inequalities of $C_k(\Lambda)$ define pairwise distinct facets. Let $E_k$ be the node--edge incidence matrix of $\Gamma_k$, and let $E'_k$ be the matrix whose columns are the stars $g_{e,c}$. Since $\Gamma_k$ is connected and bipartite, $\ker E_k^\top$ is spanned by the vector that equals $1$ on the left nodes and $-1$ on the right nodes, so $\operatorname{im}E_k$ consists of the vectors whose left and right sums agree. For each of the coarse vectors that define the $v_T$, the left sum exceeds the right sum by one. The difference of two of them therefore lies in $\operatorname{im}E_k$, and its lift lies in $\operatorname{im}E'_k$, because the sum of all copies of an edge is the lift of the corresponding column of $E_k$. Since the columns of $W_k(\Lambda)$ also include the shifts of the $v_T$ by every star, the direction space of the affine hull of $C_k(\Lambda)$ is exactly $\operatorname{im}E'_k$.

In a vector of $\ker E'_k$, the coefficients of any two copies of the same edge are equal, as one sees by comparing two rows at an endpoint whose choice functions differ only on that edge. Conversely, repeating each coefficient of a vector of $\ker E_k$ on all copies of the corresponding edge gives a vector of $\ker E'_k$. Hence $\ker E'_k\cong\ker E_k$, and $\operatorname{rank}E'_k=12(N-1)+\operatorname{rank}E_k$.

Now fix a row $F=(f,\sigma)$ and delete from $E'_k$ the $|I(f)|$ stars that are positive at $F$. A kernel vector of the remaining matrix extends by zeros to a kernel vector of $E'_k$, and by the previous paragraph it corresponds to a vector of $\ker E_k$ that vanishes on all of $I(f)$. Deleting the columns of $I(f)$ from $E_k$ lowers its rank by exactly one, because the remaining graph consists of the isolated node $f$ and a connected bipartite graph. To see the connectivity, suppose first that $f$ is a left node. Then any two remaining left nodes have a common neighbor, and every right node has a remaining left neighbor. If $f$ is a right node and $k=3$, any two left nodes still have a common neighbor. If $f$ is a right node and $k=2$, then $\Gamma_2$ is the subdivision of $K_4$, and deleting $f$ removes one subdivided edge. By the rank--nullity theorem, the matrix of the remaining stars has rank
\[
 (12N-|I(f)|)-\big[12-|I(f)|-(\operatorname{rank}E_k-1)\big]
     =\operatorname{rank}E'_k-1 .
\]
Some $v_T$ vanishes at $F$. The face $\{x_F=0\}$ of $C_k(\Lambda)$ contains this point together with its shifts by all stars that vanish at $F$, so by the rank computation it has codimension at most one. It is a proper face, because some other $v_T$ is positive at $F$. Hence every coordinate inequality $x_F\ge0$ defines a facet. These facets are pairwise distinct: the coordinate functionals restrict to the star directions as the rows of $E'_k$, which are distinct nonzero zero-one vectors and hence pairwise nonproportional. Indeed, the row of $(f,\sigma)$ determines $f$ through its set of incident edges and then $\sigma$ through the copies that it meets.

\begin{proof}[Proof of the assertions about $U_n,V_n$ in Theorem~\ref{thm:binary}]
We take the Cartesian powers $C_2(\Lambda)^n$ and $C_3(\Lambda)^n$ as base polytopes, with $\Lambda$ as in Proposition~\ref{prop:familycost}. In each of them we retain the coordinate facets, which are facets by the preceding verification applied in each block, and we wedge over all remaining facets to obtain $U_n$ and $V_n$. The two matrices of retained rows are $M_{2,n}$ and $M_{3,n}$, up to columns that are convex combinations of other columns. All generating points are integral, so the bases have rational facet descriptions, and $U_n,V_n$ are rational polytopes. By Theorem~\ref{thm:transfer}, every wedge contributes exactly two to each of $\xc(U_n)$, $\xc(V_n)$ and $\xc(U_n\times V_n)$, and these contributions cancel in the deficit. Proposition~\ref{prop:familycost} and~\eqref{eq:familyjoint} therefore give $\delta(U_n,V_n)\ge(3+32/27-4)n=5n/27$. Since the deficit is an integer, $\delta(U_n,V_n)\ge\lceil5n/27\rceil$.

For the relative assertion, note that $C_k(\Lambda)$ is the convex hull of at most $v_k=\binom{4}{k}(1+p)$ points. A facet is determined by the set of generating points that it contains, so $C_k(\Lambda)$ has at most $2^{v_k}$ facets, a bound that does not depend on $\Lambda$. Every facet of a Cartesian power comes from a facet of a single coordinate, so each base has at most $2^{v_k}n$ facets. Hence the number of wedges is $O(n)$, and by Theorem~\ref{thm:transfer} and~\eqref{eq:familyupper} the sum of the two separate extension complexities is at most $Kn$ for an absolute constant $K$. Together with the additive bound this gives the relative assertion with $\eta=5/(27K)$.
\end{proof}
The dimensions of $U_n$ and $V_n$ depend on the numbers of noncoordinate facets of the bases. We do not need to list these facets, since the construction specifies the factors as wedges over convex hulls of explicit point sets.

\input{identical_factors}

%% file: identical_factors.tex
\subsection{Identical factors by separating the alphabet sizes}\label{sec:twins}
We now prove that the deficit is unbounded also for Cartesian squares. The construction for $U_n,V_n$ uses the same four labels in every coordinate, which gives a uniform relative estimate. Here the alphabets grow from one coordinate to the next. This allows an exact counting argument that separates the cost of a new coordinate from the optimal cost of all earlier ones.

For an integer $j\ge2$ let $\mathcal F_j$ be the family of all $j$-element subsets of $[2j-1]$. A smallest hitting set of $\mathcal F_j$ has $j$ elements: the complement of any set of at most $j-1$ labels contains a member of $\mathcal F_j$, whereas any $j$ labels meet every member. Moreover, any two members of $\mathcal F_j$ intersect.

Consider the list of $t$ blocks in which block $i$ has label set $[2j_i-1]$ and family $\mathcal F_{j_i}$, and let $C_t$ be its cover number; profiles, availability and covers are defined in Section~\ref{sec:dict}. Covering each block $i$ separately by the $j_i$ singleton-block profiles of a hitting set gives $C_t\le\sum_i j_i$.

\begin{lemma}\label{lem:largealphabet}
Suppose that the first $t-1\ge1$ blocks have cover number $C$ and that the last block is $\mathcal F_j$ with $j\ge 2C$. Then $C_t=C+j$.
\end{lemma}
\begin{proof}
Consider any cover of the $t$ blocks. Let $p_0$ be the number of its profiles that specify only earlier blocks, $b_0$ the number that specify only the last block, and $m_0$ the number of \emph{mixed} profiles, which specify the last block and at least one earlier block. We use the following restriction. Fix a set $T\in\mathcal F_j$ in the last block, discard the mixed profiles whose last label is not in $T$, and erase the last block from the remaining mixed profiles. Together with the $p_0$ profiles on earlier blocks, this gives a cover of the first $t-1$ blocks: for every input of the earlier blocks and every earlier block $i$, the original cover contains a profile that specifies $i$ and is available at this input extended by $T$, and this profile survives the restriction.

Suppose first that $m_0\le j-1$. We choose $T$ to avoid the last labels of all mixed profiles. The restriction then keeps only the $p_0$ profiles on earlier blocks, so $p_0\ge C$. Moreover $b_0+m_0\ge j$, because the last labels of the profiles that specify the last block form a hitting set of $\mathcal F_j$. Hence the cover has at least $C+j$ profiles. Suppose next that $m_0\ge j-1$ and $b_0\ge1$. We choose $T$ to avoid the last labels of some $j-1$ mixed profiles. The restriction keeps at most $p_0+m_0-(j-1)$ profiles, so this number is at least $C$, and the cover again has at least $C+j$ profiles.

In the remaining case $b_0=0$. Fix an input of the earlier blocks, and call a mixed profile \emph{compatible} if all its earlier labels belong to the corresponding sets of this input. Since no profile specifies the last block alone, the last labels of the compatible mixed profiles form a hitting set of $\mathcal F_j$, so at least $j$ mixed profiles are compatible. Now choose the earlier sets independently and uniformly at random. Every mixed profile specifies at least one earlier block $i$, and its label there belongs to the random set with probability $j_i/(2j_i-1)\le2/3$. The expected number of compatible mixed profiles is therefore at most $2m_0/3$, and we obtain $m_0\ge3j/2\ge C+j$, because $j\ge2C$. In every case the cover has at least $C+j$ profiles. Conversely, a smallest cover of the earlier blocks together with $j$ singleton-block profiles in the last block is a cover of size $C+j$.
\end{proof}

\begin{lemma}\label{lem:profilebridge}
For every list of blocks $\mathcal F_{j_1},\ldots,\mathcal F_{j_t}$ as above there are a rational polytope, a matrix $M$ formed by selected facet rows of its slack matrix, and an integer $p_\star$ such that
\[
 p_\star+C_t\le\rk M\le p_\star+\sum_i j_i,\qquad
 \rk(M\ot M)\le2p_\star+\sum_i(2j_i-1).
\]
\end{lemma}
\begin{proof}
For block $i$ we repeat the construction of Section~\ref{sec:dict} with the labels $[2j_i-1]$ as left nodes and the sets in $\mathcal F_{j_i}$ as right nodes, so that $v_T$ has the value $j_i-1$ at the right node $T$. Let $D$ be the largest degree of a coarse node in any block, let $q=\max_i j_i$, and let $s_*$ be the largest number of coarse nodes in a block. We use $N_*=D+q$ copies of each coarse edge, so that the total number of stars is $p_\star=N_*\sum_i j_i\binom{2j_i-1}{j_i}$. In block $i$ let $W_i$ be the matrix with the columns $v_T$ and $v_T+\Lambda g_{e,c}$, where the common scale $\Lambda$ is specified below. The columns of $M$ are obtained by stacking one independently chosen column from each $W_i$. The identity~\eqref{eq:familychoice} holds in each block with its own sets. Using the stars and a hitting set of $j_i$ labels separately in each block gives the upper bound for $\rk M$. In $M\ot M$ we pair each block with its copy. The two sets chosen in a pair of blocks always intersect, so the $2j_i-1$ generators that equal $b_s$ in both blocks, together with the one-sided stars, suffice for this pair. This proves the upper bound for $\rk(M\ot M)$.

The lower bound for $\rk M$ follows the proof of Proposition~\ref{prop:familycost}, and we indicate the changes. Let $H=p_\star+\sum_i j_i$ and consider a nonnegative factorization of $M$ with $r\le H$ generators, normalized to have maximum entry one. The entries of a small target are at most $q$, and hence so are the coefficients in its representations. With $\tau=\Lambda^{-1/2}$, $\beta=q/(\Lambda\tau)$ and $\epsilon=q/\Lambda+H\tau$, we choose the integer scale $\Lambda$ so large that $\Lambda>q$, $\beta,\epsilon<1$ and
\[
 H(2\beta+\tau)<1,\qquad
 qHs_*\max\{\beta,\epsilon/(1-\epsilon)\}<1 .
\]
This is possible because $\tau$, $\beta$ and $\epsilon$ tend to zero as $\Lambda\to\infty$; for definiteness we take the smallest such perfect square. The probes differ from their stars by at most $q/\Lambda$, so their coefficients are less than two. A generator assigned to a star is at most $\beta<1$ outside that star, so a row where it attains its maximum belongs to every star to which it is assigned.

As before, we reserve for every nondeficient star one generator that is assigned to it alone. If $b$ stars are deficient, $a$ further generators are assigned to some star, and $u$ generators are assigned to none, then $r=p_\star-b+a+u$. A generator is assigned to at most $D$ stars, namely to copies of distinct coarse edges at one coarse node of one block. The counting of rows at a fixed endpoint shows as before that at least $N_*$ generators are assigned to each deficient star, each of them to at least two stars, so $b\le Da/N_*$. We mark stars and define good blocks as before. At most $a/N_*$ blocks are not good, since marking all copies of one coarse edge takes $N_*$ distinct generators assigned within that block. On a good block, the signed functional involves at most $s_*$ rows, vanishes on all stars and equals one at every small target. By the second inequality in the display above, the reserved and assigned generators together contribute less than one in absolute value. Thus, at every small target, every good block contains a positive entry of an unassigned generator with a positive coefficient.

We label the rows and attach profiles to the unassigned generators as in Section~\ref{sec:dict}. By the previous paragraph, the $u$ profiles cover all good blocks. For each block that is not good we add the $j_i\le q$ singleton-block profiles of a hitting set. The result is a cover of all blocks, so $C_t\le u+qa/N_*$. Consequently
\[
      r\ge p_\star+C_t+a\bigl(1-(D+q)/N_*\bigr)\ge p_\star+C_t .
\]
A factorization with more than $H$ generators also satisfies this bound, since $C_t\le\sum_i j_i$.

Finally, we take as base polytope the product of the convex hulls of the columns of the matrices $W_i$. The facet verification of Section~\ref{sec:binary} applies to these graphs as well, because each of them remains connected when a node is deleted. If a left node is deleted, every right node retains a left neighbor and any two remaining left nodes have a common neighbor. If a right node is deleted and $j_i\ge3$, any two left nodes still have a common neighbor; if $j_i=2$, the graph is a subdivided triangle, which remains connected when a right node is deleted. Hence the same kernel computation shows that every coordinate inequality defines a facet, and these facets are pairwise distinct. The rows of these facets, evaluated at the vertices of the base, form the matrix $M$ up to columns that are convex combinations of other columns. The base is rational because its vertices are integral.
\end{proof}

\begin{proof}[Proof of the assertion about $R_t$ in Theorem~\ref{thm:binary}]
Let $j_1=2$ and $j_i=2\sum_{h<i}j_h$ for $i\ge2$. A single block $\mathcal F_{j_1}$ has cover number $j_1$, and Lemma~\ref{lem:largealphabet} gives $C_t=\sum_i j_i$ by induction on $t$. By Lemma~\ref{lem:profilebridge}, $\rk M=p_\star+\sum_i j_i$ and $\rk(M\ot M)\le2p_\star+2\sum_i j_i-t$. Let $R_t$ be the polytope obtained from the base polytope of Lemma~\ref{lem:profilebridge} by wedging over all facets that are not selected. By Theorem~\ref{thm:transfer},
\[
  \delta(R_t,R_t)=2\rk M-\rk(M\ot M)\ge t .\qedhere
\]
\end{proof}
In this construction the alphabets grow rapidly with $t$, and we do not obtain a uniform relative estimate for squares; see the open questions in Section~\ref{sec:discussion}.

%% file: extension_complexity.bbl
\begin{thebibliography}{99}

\bibitem{aprile}
M. Aprile, M. Conforti, S. Fiorini, Y. Faenza, T. Huynh and M. Macchia,
\href{https://doi.org/10.1016/j.dam.2022.07.028}{Slack matrices, \(k\)-products, and 2-level polytopes},
\emph{Discrete Appl. Math.} 354 (2024), 199--213.

\bibitem{dagstuhl}
L. B. Beasley, H. Klauck, T. Lee and D. O. Theis,
\href{https://doi.org/10.4230/DagRep.3.2.127}{Communication complexity, linear optimization, and lower bounds for the nonnegative rank of matrices (Dagstuhl Seminar 13082)},
\emph{Dagstuhl Reports} 3(2) (2013), 127--143.

\bibitem{cohenrothblum}
J. E. Cohen and U. G. Rothblum,
\href{https://doi.org/10.1016/0024-3795(93)90224-C}{Nonnegative ranks, decompositions, and factorizations of nonnegative matrices},
\emph{Linear Algebra Appl.} 190 (1993), 149--168.

\bibitem{ccz}
M. Conforti, G. Cornu\'ejols and G. Zambelli,
\href{https://doi.org/10.1007/s10479-012-1269-0}{Extended formulations in combinatorial optimization},
\emph{Ann. Oper. Res.} 204 (2013), 97--143.

\bibitem{grt}
J. Gouveia, R. Z. Robinson and R. R. Thomas,
\href{https://doi.org/10.1007/s00454-013-9533-x}{Polytopes of minimum positive semidefinite rank},
\emph{Discrete Comput. Geom.} 50 (2013), 679--699.

\bibitem{gps}
F. Grande, A. Padrol and R. Sanyal,
\href{https://doi.org/10.1007/s00454-017-9925-4}{Extension complexity and realization spaces of hypersimplices},
\emph{Discrete Comput. Geom.} 59 (2018), 621--642.

\bibitem{kaibel}
V. Kaibel, \href{https://ufdcimages.uflib.ufl.edu/UF/00/09/00/46/00085/optima85.pdf}{Extended formulations in combinatorial optimization},
\emph{Optima} 85 (2011), 2--7.

\bibitem{ksz}
M. Kwan, L. Sauermann and Y. Zhao,
\href{https://doi.org/10.1090/tran/8614}{Extension complexity of low-dimensional polytopes},
\emph{Trans. Amer. Math. Soc.} 375 (2022), 4209--4250.

\bibitem{padrol}
A. Padrol,
\href{https://doi.org/10.1137/16M1063721}{Extension complexity of polytopes with few vertices or facets},
\emph{SIAM J. Discrete Math.} 30 (2016), 2162--2176.

\bibitem{shitov}
Y. Shitov, \href{https://doi.org/10.4310/ACTA.2019.v222.n2.a3}{Counterexamples to Strassen's direct sum conjecture},
\emph{Acta Math.} 222 (2019), 363--379.

\bibitem{shitovoverview}
Y. Shitov, \href{https://www.researchgate.net/publication/417554755}{Illusions of stolen science. II. Mathematicians}, ResearchGate preprint, publication no.~417554755, 5~November 2026. Section~6.1 reports the result. The proof is in the AI-generated manuscript (OpenAI GPT-6 Astra Pro) \emph{Extension complexity can be strictly subadditive under Cartesian products}, dated 2~November 2026, which is article~6.1 of the archive of 24 manuscripts distributed with the preprint (link in its Section~4).

\bibitem{twz}
H. R. Tiwary, S. Weltge and R. Zenklusen,
\href{https://doi.org/10.1016/j.ipl.2017.07.011}{Extension complexities of Cartesian products involving a pyramid},
\emph{Inform. Process. Lett.} 128 (2017), 11--13.

\bibitem{weltge}
S. Weltge, \href{https://doi.org/10.25673/4350}{\emph{Sizes of Linear Descriptions in Combinatorial Optimization}},
doctoral dissertation, Otto-von-Guericke-Universit\"at Magdeburg, 2015.

\bibitem{cargese}
S. Weltge, Extension complexity and Cartesian products, in
\href{https://www.cargese.org/2022/open-problems.pdf}{\emph{Carg\`ese Workshop on Combinatorial Optimization 2022, Open Problems}}, 2022.

\bibitem{yannakakis}
M. Yannakakis, \href{https://doi.org/10.1016/0022-0000(91)90024-Y}{Expressing combinatorial optimization problems by linear programs},
\emph{J. Comput. System Sci.} 43 (1991), 441--466.

\end{thebibliography}
